\documentclass[11pt,letterpaper]{article}
\pdfoutput=1
\usepackage{jheppub}

\usepackage[utf8]{inputenc} % utf8 input good for non-ascii characters
\usepackage[T1]{fontenc} % Good for hyphenating non-ascii characters
\usepackage{lmodern}
\usepackage{graphicx}

\usepackage{booktabs} % Better tables
\usepackage{array} % Allows us to define custom column types for tables
\newcolumntype{L}{>{$}l<{$}} % A left aligned maths column type

\usepackage{microtype} % Improves typesetting
\usepackage[margin=7mm]{caption} % Format captions
\usepackage[font=small,labelfont=bf]{caption}
\usepackage{xfrac} % Nice small fractions
\makeatletter
\g@addto@macro\@floatboxreset\centering
\makeatother

\usepackage[italic]{hepnames} % Add particle name macros (e.g. \PBs)
\renewcommand{\PBs}{\HepParticle{\PB}{\Pqs}{}\xspace}

\renewcommand{\PBd}{\HepParticle{\PB}{\Pqd}{}\xspace}
\usepackage{braket} % Adds Dirac bra-ket notation
\usepackage{siunitx} % Add support for units
\DeclareSIUnit\fb{\femto\barn}
\DeclareSIUnit\invps{\ps^{-1}}
\usepackage{suffix} % Allows us to define a starred command, for CKM macro later

\newcommand{\BigO}[1]{\ensuremath{\mathcal{O}\left(#1\right)}}

\newcommand{\tauBs}{\tau(\PBs)}
\newcommand{\tauBd}{\tau(\PBd)}

\newcommand{\Lagrangian}{\ensuremath{\mathcal{L}}\xspace}
\newcommand{\hc}{\text{h.c.}}

\newcommand{\lambdaHS}{\lambda_{HS}}
\newcommand{\lambdaHSprime}{\lambda'_{HS}}
\newcommand{\lambdaPhiS}{\lambda_{\Phi S}}

\newcommand{\V}[1]{\ensuremath{V_{#1}^{}}} 
\WithSuffix\newcommand\V*[1]{\ensuremath{V_{#1}^*}}

\newcommand{\be}{\begin{equation}}
\newcommand{\ee}{\end{equation}}

\title{Anomalies and accidental symmetries: charging the scalar leptoquark under $L_\mu - L_\tau$}

\author[a]{Joe Davighi,}
\author[b,c]{Matthew Kirk,}
\author[b,c]{and Marco Nardecchia}
\affiliation[a]{DAMTP, University of Cambridge, Wilberforce Road, Cambridge, 
CB3 0WA, United Kingdom}
\affiliation[b]{Sapienza - Universit\'a di Roma, Piazzale Aldo Moro 2, 00185, Roma, Italy}
\affiliation[c]{INFN - Sezione di Roma, Piazzale Aldo Moro 2, 00185, Roma, Italy}

\emailAdd{jed60@cam.ac.uk}
\emailAdd{matthew.kirk@roma1.infn.it}
\emailAdd{marco.nardecchia@cern.ch}

\abstract{While the $S_3$ scalar leptoquark presents a possible tree-level explanation of the $b \to s \ell \ell$ flavour anomalies, it suffers from two conceptual problems which are often disregarded by model-builders. Firstly, the quantum numbers of the $S_3$ allow for a renormalisable diquark operator that would trigger rapid proton decay unless its coupling were tuned away. Secondly, one expects the leptoquark to have generic couplings to leptons, which require tuning to avoid stringent experimental bounds on lepton flavour violation. By gauging a $U(1)$ current that acts as $L_\mu - L_\tau$ on the Standard Model (SM) fermions, and under which the leptoquark has charge $-1$, one can remedy both these problems. The additional $U(1)$, which is spontaneously broken at some high scale, is associated with a massive $Z^\prime$ gauge boson and a scalar SM singlet $\Phi$, which play no direct role in mediating the anomalous $B$ meson decays. By computing one- and two-loop mass corrections, we show that this pair of particles can be hidden away at much higher mass scales without destabilising either the Higgs or the leptoquark masses. The only low-energy relic of gauging $L_\mu - L_\tau$ is thus the accidental global symmetry structure of the lagrangian. On the other hand, we find quite generally that an $S_3$ leptoquark that mediates the $b \to s \ell \ell$ anomalies cannot be much heavier than a few TeV without itself inducing large Higgs mass corrections.
}

\begin{document} 
\maketitle
\flushbottom

\section{Introduction} \label{intro}

It is well known that the collection of anomalies recently observed in neutral current decays of $B$ mesons~\cite{Aaij:2017vbb,Aaij:2019wad,Aaboud:2018mst,Chatrchyan:2013bka,CMS:2014xfa,Aaij:2017vad,Aaij:2013qta,Aaij:2015oid,Sirunyan:2017dhj,Aaboud:2018krd} can be explained at tree level by either a leptoquark or a neutral heavy gauge boson, whose mass over coupling is around 30--40 TeV. After integrating out either of these heavy states and matching onto the weak effective theory (WET) one can produce four fermion operators of the form $bs\ell\ell$, $\ell \in \{e,\mu\}$, which can mediate the anomalies. Of the various chirality combinations, global fits to the $B$-anomaly data favour a non-zero Wilson coefficient for the particular WET operator~\cite{Aebischer:2019mlg,Alguero:2019ptt}
\begin{equation}
  \mathcal{O}_L = (\overline{s_L} \gamma_\rho b_L) 
  (\overline{\mu_L} \gamma^\rho \mu_L), \label{eq:WET operator}
\end{equation}
{\em i.e.} with left-handed quark and left-handed muon currents.
Indeed, turning on the single Wilson coefficient for this operator provides an excellent global fit to the data with a pull of 6.6$\sigma$ relative to the Standard Model (SM), according to Ref.~\cite{Aebischer:2019mlg}.\footnote{In light of an update to LHCb's angular analysis of the $B^0 \to K^{\ast 0}\mu^+ \mu^-$ decay~\cite{Aaij:2020nrf}, the statistical pull of this new physics scenario with respect to the SM has increased slightly. } 

Of the various leptoquarks that can mediate $b s \ell \ell$ processes, only a handful have the right quantum numbers to couple to left-handed quarks and lepton doublets, and so contribute to the operator $\mathcal{O}_L$. Two of these are Lorentz vector states, often called the $U_1$ and $U_3$ leptoquarks, which transform in the $(\mathbf{3},\mathbf{1})_{2/3}$ and $(\mathbf{3},\mathbf{3})_{2/3}$ representations of $SU(3)\times SU(2)_L \times U(1)_Y$. A theory in which the SM is supplemented by one of these vector leptoquarks is on its own non-renormalisable, and so must be embedded in an ultraviolet (UV) completion. A variety of suitable UV completions, in which the vector leptoquark typically emerges as a massive gauge boson from some spontaneously broken non-abelian extension of the SM gauge symmetry, have been developed in great detail in recent years~\cite{Assad:2017iib,DiLuzio:2017vat,Bordone:2017bld, Calibbi:2017qbu,Barbieri:2017tuq,Greljo:2018tuh}. 

\begin{figure}
\includegraphics[width=0.35\textwidth]{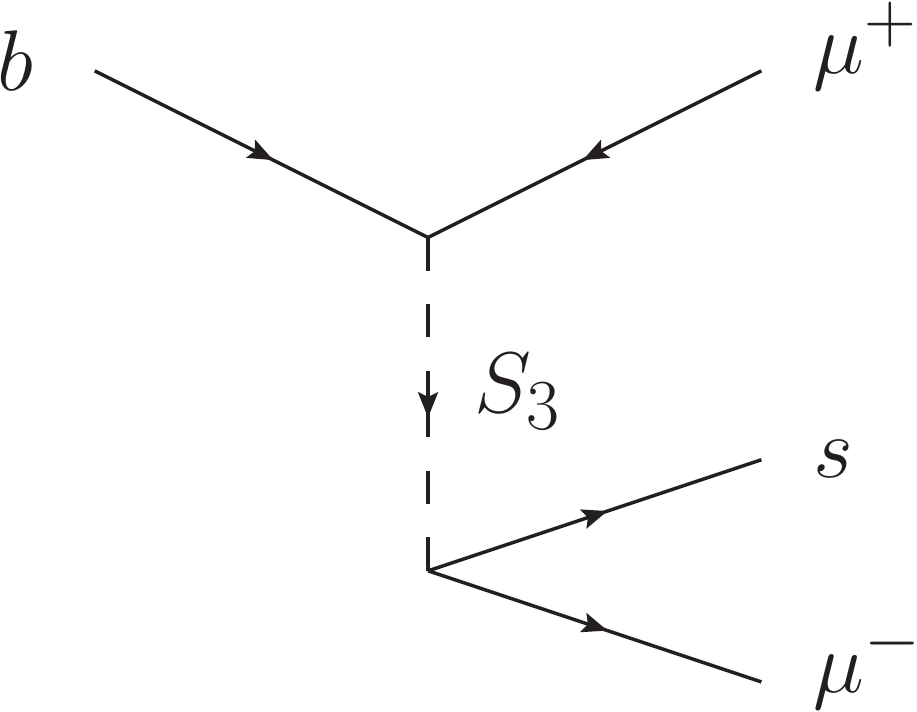}
\caption{Our model features an $S_3$ scalar leptoquark that is charged under $L_\mu - L_\tau$, which therefore couples only to second family leptons. This leptoquark mediates the anomalous $b\to s \ell\ell$ transitions via a tree-level contribution that induces $\Delta C_9^\mu = -\Delta C_{10}^\mu$. }
\label{fig:CL}
\end{figure}

If one wishes to avoid this task of building a compelling UV completion, one might prefer to consider scalar leptoquarks, since extending the SM by a scalar leptoquark gives a consistent theory on its own (from the point of view of renormalisability -- there remains the separate challenge of explaining why such a scalar should be so light compared to the Planck scale).\footnote{One attractive possibility is that the scalar leptoquark is light because it is a pseudo Nambu Goldstone boson (pNGB) associated with global symmetry breaking in a new strong sector, possibly alongside the Higgs boson within the framework of partial compositeness~\cite{Gripaios:2009dq,Gripaios:2014tna}.} Only one scalar leptoquark, the one dubbed $S_3$ that transforms in the $(\bar{\mathbf{3}},\mathbf{3})_{1/3}$ representation, has couplings to the left-handed quark and lepton doublets of the SM, via terms
\begin{equation}
\mathcal{L} \supset \lambda^{QL}_{ij} \overline{Q^c_i} (i \sigma_2 \sigma_a) L_j S_3^a \label{eq:LQ-quark-lepton_generic}
\end{equation}
in the lagrangian (where $i,j$ are family indices and $a$ is an adjoint $SU(2)$ index,  and we assume the down-aligned mass basis for quarks). Only this scalar leptoquark can thence give rise to the WET operator~(\ref{eq:WET operator}) after being integrating out (see Fig.~\ref{fig:CL}), provided $\lambda^{QL}_{22}$ and $\lambda^{QL}_{32}$ are non-zero, which has the preferred chiral structure to explain the anomalous measurements in the $bs\mu\mu$ system (as was first proposed in~\cite{Hiller:2014yaa}).

Extensions of the SM by an $S_3$ leptoquark suffer from two conceptual difficulties, which are often swept under the rug. Firstly, the quantum numbers of the $S_3$ leptoquark allow for a gauge-invariant `diquark' operator in the lagrangian, schematically 
\begin{equation}
\mathcal{L} \supset \lambda^{QQ}_{ij} \overline{Q_i^c} S_3 Q_j. \label{eq:diquark}
\end{equation}
Operators of this type violate baryon number, and in particular lead to rapid proton decay (especially the $\lambda^{QQ}_{11}$ component). In order to pass the stringent experimental constraints on the proton lifetime~\cite{Tanabashi:2018oca}, this diquark term is often set to zero in an {\em ad hoc} fashion (see {\em e.g.} the phenomenological studies in~\cite{Angelescu:2018tyl,Allanach:2019zfr}). A more elegant way to outlaw the  diquark operator is to embed the $S_3$ leptoquark in representations of a larger unified gauge group, such as the $\mathbf{126}$-dimensional representation of $SO(10)$, in which the diquark operator is forbidden by the $SO(10)$ symmetry~\cite{Senjanovic:1982ex,Dorsner:2016wpm}. The same can be achieved in an $SU(5)$ GUT~\cite{Dorsner:2017wwn,Dorsner:2017ufx}.

The second troublesome feature of $S_3$ leptoquark models concerns their couplings to the lepton sector. Based on the symmetries of the theory, one naturally expects the $S_3$ leptoquark to couple to all three generations of left-handed lepton doublets, via the lagrangian~(\ref{eq:LQ-quark-lepton_generic}). But a generic matrix $\lambda^{QL}_{ij}$ of couplings is dangerous from the point of view of phenomenology, because it leads to sizeable contributions to a number of rare lepton flavour violating (LFV) processes. For example, decays $\ell \rightarrow \ell' \gamma$ are induced at loop level (with the $\mu \rightarrow e \gamma$ branching ratio being especially tightly constrained), as are decays such as $Z \rightarrow \ell \ell'$ and $\ell \rightarrow \ell' \ell' \ell''$. Other LFV processes, such as $B\rightarrow K\mu\tau$, would be induced even at tree-level. The experimental bounds are often strongest in LFV processes involving electrons; for example, the branching ratio $B(\mu\rightarrow 3e) < 1.0 \times 10^{-12}$ at 90\% C.L.~\cite{Bellgardt:1987du}. As was the case for the diquark operator discussed above, the couplings of an $S_3$ leptoquark to electrons must be tuned close to zero by hand in order to pass these pressing constraints from LFV. Even with such aggressive assumptions, LFV measurements in the $\mu-\tau$ system place important bounds on the leptoquark couplings, as discussed for example in~\cite{Dorsner:2017ufx}. Note that embedding the $S_3$ in a grand unified theory based on $SO(10)$ or $SU(5)$, along the lines described in the previous paragraph, does nothing to ameliorate this problem.

In this paper, we suggest a simple mechanism for remedying both of these problems within the context of $S_3$ models for the neutral current $B$-anomalies. If the $S_3$ leptoquark is charged under an additional $U(1)$ symmetry, under which the SM fields are assigned appropriate charges, then both the diquark operator (\ref{eq:diquark}) and the leptoquark couplings to electrons and tauons will be banned at the renormalisable level. In other words, one can achieve the desired structure of leptoquark couplings,
\begin{equation}\label{eq:coupling structure}
\lambda^{QL}_{ij} = \alpha_i \delta_{j2}, \quad \lambda^{QQ}_{ij} = 0
\end{equation}
with a particular choice of $U(1)$ charges, which we soon discuss, where the $\alpha_i$ are three {\em a priori} complex couplings.
This $U(1)$ symmetry will be spontaneously broken, most likely by some scalar SM singlet field $\Phi$ acquiring a vev at some high energy scale above the electroweak scale, leading to a massive $Z^\prime$ gauge boson.
(However, it is important to stress that this $Z^\prime$ boson plays no role in mediating the $B$ anomalies, which are mediated solely by the $S_3$ leptoquark.)
In order for the theory to be self-consistent, the $U(1)$ charge assignment must be anomaly-free. The space of family-dependent anomaly-free $U(1)$ extensions of the SM gauge symmetry has been well-studied recently~\cite{Allanach:2018vjg,Allanach:2019uuu,Allanach:2020zna,Costa:2020krs}, and is unhelpfully vast unless further criteria are imposed.

\subsection*{The gauge sector extension}

We here explore an especially well-motivated choice, which is to gauge the $L_\mu - L_\tau$ current, under which we assign the $S_3$ leptoquark a charge of $-1$. This choice satisfies our desired criteria.
Firstly, since all quarks are uncharged under the $U(1)$ while the leptoquark is charged, the diquark operators (\ref{eq:diquark}) cannot be $U(1)$ invariant, and thus baryon number is restored at the renormalisable level.\footnote{If our goal is to ban the $B$-violating diquark operator using a gauge symmetry, then charging the $S_3$ LQ under $U(1)_{L_\mu-L_\tau}$ is arguably a more minimal extension of the SM than embedding the $S_3$ leptoquark in an $\mathbf{126}$-dimensional ($\mathbf{45}$-dimensional) representation of a unified $SO(10)$ ($SU(5)$) theory. }  Secondly, of the couplings in~(\ref{eq:LQ-quark-lepton_generic}), only the muonic contributions ($\lambda^{QL}_{i2}$) are $U(1)$ invariant and permitted in the renormalisable lagrangian, thereby picking out the muon direction needed to mediate the anomalous $B$-decays, while simultaneously suppressing LFV processes induced by the leptoquark. Thirdly, the $L_\mu - L_\tau$ charge assignment is anomaly-free. We remark that a similar idea was explored in~\cite{Hambye:2017qix}, in which proton decay was prohibited by imposing a global \(L_\mu + 2 L_e - 3 L_\tau\) symmetry (which can be made anomaly-free by including three right-handed neutrinos).

The particular choice of $U(1)$ that we gauge has additional virtues. By gauging what is otherwise an accidental symmetry of the SM we allow for a generic Yukawa sector for the quarks at the renormalisable level, as well as a strictly diagonal Yukawa matrix for the charged leptons. Indeed, if we make a renormalisable Yukawa sector a requirement, then, given only the SM chiral fermion content, the most general anomaly-free $U(1)_X$ charge assignment is of the form $L_i - L_j+aY$ where $Y$ denotes hypercharge, for any pair of lepton numbers $L_i$ and $L_j$ and any rational number $a$.\footnote{We here choose to set $a=0$ for simplicity, so that the $Z^\prime$ has no tree-level couplings to quarks (such couplings will, however, be generated radiatively). } As discussed in~\cite{Altmannshofer:2019xda}, a strictly diagonal charged lepton Yukawa matrix implies that the charged lepton mass eigenstates are aligned with the weak eigenstates. Hence, even though this theory features a $Z^\prime$ boson with non-universal couplings to charged leptons, there is no danger of LFV being induced by the $Z^\prime$. 

Taking a more general view, by gauging an accidental symmetry of the SM we ensure that the global part of this symmetry remains an accidental symmetry of our BSM theory, thereby guaranteeing safety from the LFV constraints. Our leptoquark model shares these virtues with a number of $Z^\prime$ models for the $b\to s \ell\ell$ anomalies in which the $Z^\prime$ also arises from gauging, say, $L_\mu - L_\tau$ symmetry~\cite{Altmannshofer:2014cfa,Altmannshofer:2015mqa,Altmannshofer:2016jzy}, or indeed a more general linear combination of the SM's accidental symmetries~\cite{Altmannshofer:2019xda}. By way of comparison, recall that in these $Z^\prime$ models the required flavour-violating couplings to down-type quarks are absent at tree level; rather, they are typically generated by introducing a fourth family of vector-like quarks with somewhat {\em ad hoc} couplings to the SM fields. This paper explores a different avenue in which we dispense totally with the fourth family of heavy quarks, at the expense of introducing only one extra state, the $S_3$ leptoquark.\footnote{Another noteworthy distinction is that our $(L_\mu-L_\tau)$-charged leptoquark model explains the $B$-anomaly data via the left-handed operator $\mathcal{O}_L$ of (\ref{eq:WET operator}), whereas models in which the corresponding $L_\mu-L_\tau$ heavy gauge boson mediates the anomalies do so via the vector-like operator $(\overline{s_L}\gamma_\rho b_L)(\overline{\mu}\gamma^\rho \mu)$. The global fits of Ref.~\cite{Aebischer:2019mlg} favour the former scenario, largely because the latter does not give a BSM contribution to $BR(B_s \to \mu^+\mu^-)$ in which a $\sim 2\sigma$ discrepancy with the SM persists~\cite{Aaboud:2018mst,Chatrchyan:2013bka,CMS:2014xfa,Aaij:2017vad}. } The role of the $Z^\prime$ here is only to protect the accidental symmetries of the SM.

\subsection*{Finite naturalness of leptoquark models for the $B$ anomalies}

In this setup, the $Z^\prime$ and $\Phi$ particles associated with the extended gauge sector play no role in mediating the $B$-physics anomalies, and so their masses and couplings are not tied to the energy scale probed by the anomaly. It would appear that the $Z^\prime$ is phenomenologically inert, and that its effects can be decoupled if we wish, by taking the scale $v_\Phi$ of $U(1)_X$ breaking to very high energies, or by taking the gauge coupling $g_X$ to be very small. Things are not quite so simple, at least {\em a priori}, because taking $v_\Phi$ arbitrarily large might destabilise both the Higgs mass and the mass of the $S_3$ leptoquark through loop corrections. Nonetheless, in \S \ref{sec:naturalness} we compute finite loop corrections to the Higgs and leptoquark masses, and find that there exists a limit in which the $Z^\prime$ and $\Phi$ masses can be taken parametrically larger than the other mass scales in the theory, while their loop corrections to the Higgs and leptoquark masses remain parametrically small. In other words, we can hide away the particles associated with the extended gauge sector at very high energies, with their only phenomenological signature being the particular symmetry structure (\ref{eq:coupling structure}) of the low-energy lagrangian.

Perhaps more importantly, we also compute loop corrections to the Higgs mass from loops involving the $S_3$ leptoquark. Because the leptoquark mediates the $b \to s \ell \ell$ anomalies, these loop corrections cannot be tuned away, and their size is unavoidably tied to the magnitude of the $b\to s \ell\ell$ anomalies. We identify the parameter space of $S_3$ leptoquark models (quite generally) that is finitely natural~\cite{Farina:2013mla}, {\em i.e.} for which a mass hierarchy $M_{S_3} \gg M_H$ is radiatively stable. For example, if one prefers a $U(2)$-like flavour structure in which the leptoquark couples predominantly to the third family, then a finitely natural hierarchy can only be maintained  for leptoquark masses $M_{S_3} \lesssim 5$ TeV or so.
Moreover, an unavoidable electroweak correction means naturalness can only be maintained for \(M_{S_3} \lesssim 5.8\,\text{TeV}\), irrespective of the flavour structure.

A discussion of naturalness, and its interplay with the scale of new physics suggested by the $B$ anomaly data, has thus far been lacking in the model-building literature addressing the $B$-physics anomalies. In this paper, we find that computing the Higgs mass corrections, and requiring that they are not too big, leads to interesting hints about the mass and couplings of an $S_3$ leptoquark that can address the $B$ anomalies. In a future work, we will explore in more detail (and in greater generality) the implications of naturalness for models that explain the $b\to s \ell\ell$ anomalies.

The structure of the rest of the paper is as follows. In \S \ref{sec:setup} we set out the charged leptoquark model. After unpacking the various terms in the lagrangian, we discuss the issue of proton stability in \S \ref{sec:proton}, and then describe the $Z^\prime$ sector of the model in \S \ref{sec:Zprime}. We devote \S \ref{sec:naturalness} to considerations of finite naturalness, whereby which we establish that a mass hierarchy $M_{Z^\prime} \gg M_{S_3} \gg M_H$ is stable to finite loop corrections, and moreover that a limit exists in which the $Z^\prime$ essentially decouples from the phenomenology. In \S \ref{sec:phenomenology} we explore the phenomenology of the model in this ``decoupling limit'', before concluding.

\section{Charging the $S_3$ leptoquark} \label{sec:setup}

We consider an extension of the SM by a $U(1)_X$ gauge symmetry, where $X=L_\mu - L_\tau$ for the SM fields. The $U(1)_X$ gauge symmetry  will  be  spontaneously  broken  by  the  vacuum expectation value (vev) $v_\Phi$ of a SM singlet scalar field $\Phi$ that is charged under $U(1)_X$, leading to a massive $Z^\prime$ gauge boson. Finally, we suppose there is an $S_3$ leptoquark with quantum numbers $(\bar{\mathbf{3}},\mathbf{3})_{1/3}$ under the SM gauge symmetry, and with a charge of $-1$ under $U(1)_X$. This completes the field content of our model, which is summarised in Table~\ref{tab:bsm_field_content}.

\begin{table}
\begin{tabular}{@{}LLLLL@{}}
\toprule
\text{Field} & SU(3)_c & SU(2)_L & U(1)_Y & U(1)_X\\
\midrule
Q_L & \mathbf{3} & \mathbf{2} & \sfrac{1}{6} & 0\\
\Pqu_R & \mathbf{3} & \mathbf{1} & \sfrac{2}{3} & 0\\
\Pqd_R & \mathbf{3} & \mathbf{1} & -\sfrac{1}{3} & 0\\
L_L^i & \mathbf{1} & \mathbf{2} & -\sfrac{1}{2} & \delta_{i2}-\delta_{i3} \\
\Pe_R^i & \mathbf{1} & \mathbf{1} & -1 & \delta_{i2}-\delta_{i3} \\
\PH & \mathbf{1} & \mathbf{2} & \sfrac{1}{2} & 0\\
\hline
\hline
S_3 & \mathbf{\bar{3}} & \mathbf{3} & \sfrac{1}{3} & -1 \\
\Phi & \mathbf{1} & \mathbf{1} & 0 & \hat{Q}_\Phi \\
%\hline
%\hline
N_R^i & \mathbf{1} & \mathbf{1} & 0 & \delta_{i2}-\delta_{i3} \\
\bottomrule
\end{tabular}
\caption{The field content of the charged leptoquark model. In addition to the SM fields, there is a $U(1)_X$ gauge field with flavour non-universal couplings to SM leptons, as well as an $S_3$ scalar leptoquark and a SM singlet $\Phi$. Finally, including three right-handed neutrino fields $N_R^i$ facilitates a description of neutrino masses and mixing angles, which we comment on briefly in \S \ref{sec:proton}. }
\label{tab:bsm_field_content}
\end{table}

The renormalisable lagrangian contains the following terms
\begin{align}
\Lagrangian &= \Lagrangian_{\text{SM}}+[\alpha^i \overline{Q^c_i} (i \sigma_2 \sigma_a) L_2 S_3^a +\hc] \nonumber \\
&+ |D_\mu S_3|^2 - M_{S_3}^2 |S_3|^2 -\lambda_S |S_3|^4 - \lambda_{HS} |H|^2 |S_3|^2 - \lambda'_{HS} |H^\dagger \sigma_a S_3^a|^2 \nonumber \\
&+ |D_\mu \Phi|^2 + \mu_\Phi^2 |\Phi|^2 - \lambda_\Phi |\Phi|^4 - \lambda_{\Phi S} |\Phi|^2 |S_3|^2 - \lambda_{\Phi H} |H|^2 |\Phi|^2 \nonumber \\
&- \frac{1}{4} X_{\mu\nu} X^{\mu\nu} - \frac{\sin \epsilon}{2} B_{\mu\nu} X^{\mu\nu}- g_X X_\mu \sum_\psi \hat{Q}_\psi \bar{\psi} \gamma^\mu \psi \label{eq:LQ-quark-lepton}  \,,
\end{align}
where $Q_i =(V_{ji}^\ast u^j_L, d^i_L)^T$ and $L_i = (U_{ji}^\ast\nu^j_L, \ell^i_L)^T$ denote the quark and lepton $SU(2)_L$ doublets, and $V$ and $U$ are the CKM and PMNS matrices respectively.
The first line of Eq.~(\ref{eq:LQ-quark-lepton}) contains all the terms from the SM lagrangian, and the all-important leptoquark coupling to quark-muon pairs, which we shall elaborate on shortly. The notation $Q^c=C\overline{Q}^T$ denotes charge conjugation, where as usual $\overline{\psi}=\psi^\dagger \gamma^0$ and the charge conjugation matrix is $C=i\gamma^2\gamma^0$. 

The second and third lines of Eq.~(\ref{eq:LQ-quark-lepton}) contain the various kinetic and potential terms from the extended scalar sector of the theory, specified by the real parameters $M_{S_3}$, $\mu_\Phi$ (both mass dimension one), $\lambda_{HS}$, $\lambda^{\prime}_{HS}$, $\lambda_\Phi$, $\lambda_{\Phi S}$, and $\lambda_{\Phi H}$ (all dimensionless). Note that the sign of the quadratic term for $S_3$ implies that its vev is at $S_3=0$. Finally, the fourth line contains terms associated with the new gauge boson $X_\mu$ corresponding to the gauging of $U(1)_X$, where $X_{\mu\nu}=\partial_{[\mu} X_{\nu]}$, $g_X$ is the gauge coupling for $U(1)_X$, and $\hat{Q}_\psi$ denotes the charge of a SM fermion $\psi$ under $U(1)_X$, as given in Table~\ref{tab:bsm_field_content}. The second term on the fourth line is a tree-level contribution to kinetic mixing between the new $U(1)_X$ gauge boson and $B_\mu$, the gauge field for hypercharge. We shall discuss the effects of this kinetic mixing in \S \ref{sec:Zprime}.

In order to unpack the leptoquark couplings in Eq.~(\ref{eq:LQ-quark-lepton}), it is helpful to define linear combinations of the components $S_3^a$ which are eigenstates of the electric charge operator, {\em viz.}
\begin{equation}
S_3^{4/3} = \frac{S_3^1 - i S_3^2}{\sqrt{2}} \,, \qquad S_3^{1/3} = S_3^3 \,, \qquad S_3^{-2/3} = \frac{S_3^1 + i S_3^2}{\sqrt{2}}, \label{eq:ladder basis}
\end{equation}
where the superscript $4/3$, $1/3$, and $-2/3$ denote electric charges.
Then, after expanding the quark and lepton $SU(2)$ doublets, Eq.~(\ref{eq:LQ-quark-lepton}) decomposes into the terms
\begin{equation}
\Lagrangian \supset \begin{aligned}[t]
&-\alpha^i U_{j2}^\ast S_3^{1/3} \overline{d^c_i} P_L \nu^j - \sqrt{2} \alpha^i S_3^{4/3} \overline{d^c_i} P_L \mu \\
&+ \sqrt{2} \alpha^k V^\ast_{ik} U_{j2}^\ast S_3^{-2/3} \overline{u^c_i} P_L \nu^j - \alpha^k V^\ast_{ik} S_3^{1/3} \overline{u^c_i} P_L \mu + \hc ,
\end{aligned}
\end{equation}
The second term on the right-hand-side gives the leptoquark coupling to muons and down-type quarks, which shall be responsible for mediating the anomalous $B$ meson decays, which we address in \S \ref{sec:B anomalies}.

Importantly, quark flavour violation is linear~\cite{Gripaios:2015gra} in our model,\footnote{Consequently, the charged leptoquark framework (as for any single leptoquark model) is also an example of the more general flavour structure known as `rank-one flavour violation', explored recently in~\cite{Gherardi:2019zil}.}  with new physics coupling to the particular direction in $SU(3)$ quark flavour space
\begin{equation}
\tilde{Q} \equiv \alpha^i Q_i,
\end{equation}
where $\alpha_i\in \mathbb{C}^3$ is a 3-component complex vector. This vector could be naturally aligned in some sense with the SM flavour structure, with a plausible ansatz being that $(\alpha_1, \alpha_2, \alpha_3) \propto (V_{ub}, V_{cb}, V_{tb})$ up to order one numbers. This is what one expects in several well-motivated cases, such as the case where the $S_3$ leptoquark is embedded in a partial compositeness framework~\cite{Kaplan:1991dc,KerenZur:2012fr,Gripaios:2014tna} (which, as we have already suggested, would offer an attractive explanation for the scalar leptoquark being light). Such an ansatz also follows from the minimal flavour violation (MFV) hypothesis~\cite{DAmbrosio:2002vsn,Barbieri:2011ci}, or more generally from imposing a $U(2)^n$ flavour symmetry on the new physics (see {\em e.g.} ~\cite{Barbieri:2011ci,Barbieri:2015yvd,Buttazzo:2017ixm}). In what follows, we shall refer to such flavour structures in the quark sector as being ``$U(2)$-like'', in general.

As we discussed in the Introduction, the lepton sector is very clean.
The alignment of new physics with the muon is guaranteed by charging the leptoquark under a lepton family-non-universal gauge symmetry, here $X=L_\mu - L_\tau$ (on the SM fields). This offers a natural, symmetry-based explanation for the observation of lepton flavour universality violation in the measurements of $R_K$ and $R_{K^\ast}$, in particular the preferred alignment of new physics with the muon direction, while also preventing dangerous new physics contributions to lepton flavour violation.

At this stage, keeping all the terms generically allowed in the lagrangian (\ref{eq:LQ-quark-lepton}), we have introduced 14 new parameters:
\begin{equation}
\{\alpha_i, g_X, M_{S_3}, \lambda_S, \lambdaHS, \lambdaHSprime, \mu_\Phi, \lambda_\Phi, \lambdaPhiS, \lambda_{\Phi H}, \epsilon, \hat{Q}_\Phi\} \, .
\end{equation}
The mass parameter \(\mu_\Phi\) is related to the other parameters \(v_\Phi\), \(M_X\) and \(M_\Phi\) as
\begin{equation}
\mu_\Phi = \sqrt{\lambda_\Phi} v_\Phi = \frac{\sqrt{\lambda_\Phi} M_X}{g_X \hat{Q}_\Phi} = \frac{M_\Phi}{\sqrt{2}}.
\end{equation}
Hence we can eliminate \(\{\mu_\Phi, \lambda_\Phi\}\) and write everything in terms of the equivalent set
\begin{equation}\label{eq:parameters}
\{\alpha_i, g_X, M_{S_3}, M_\Phi, M_X, \lambda_S, \lambdaHS, \lambdaHSprime, \lambdaPhiS, \lambda_{\Phi H}, \epsilon, \hat{Q}_\Phi\}.
\end{equation}
Eventually, we will see how this set of parameters can be trimmed down substantially when we come to discuss phenomenology.

\subsection{Proton stability} \label{sec:proton}

One of our main motivations for charging the $S_3$ leptoquark under a $U(1)_X$ gauge symmetry is to preserve baryon number (most pressingly, to prevent proton decay), which is an accidental symmetry of the SM. If the leptoquark had no $U(1)_X$ charge, then nothing forbids one from writing down the renormalisable `diquark' operators
\begin{equation}
\mathcal{L}\supset \lambda^{QQ}_{ij}\;  \overline{Q^c_i} (i \sigma_2 \sigma_a S_3^a)^\dagger Q_j, \label{eq:diquark 2}
\end{equation}
where $\sigma_a$ are the Pauli matrices, and $Q^c$ denotes charge conjugation as defined above. These operators violates baryon number, and would facilitate proton decay by direct $s$-channel leptoquark exchange. Because the leptoquark in our theory is charged under $U(1)_X$, while all quark fields are neutral, the operator \ref{eq:diquark 2} clearly carries net $U(1)_X$ charge and so is banned by the $U(1)_X$ gauge symmetry. 

To achieve proton stability, however, it does not suffice to simply check whether the diquark operators are present or not; baryon number ($B$) can also be violated due to terms in the scalar potential if there are new scalars that couple to SM fermion bilinears. For an example, consider the $\tilde{R}_2$ scalar leptoquark, which has SM quantum numbers $(\bar{\mathbf{3}},\mathbf{2})_{1/6}$. Even though there is no diquark operator, there is a quartic coupling $\tilde{R}_2 \tilde{R}_2 \tilde{R}_2 H$ which, after integrating out $\tilde{R}_2$ and expanding the Higgs around its vev, leads to the decay $p\rightarrow \pi^+ \pi^+ e^- \nu\nu$ at tree-level, and thus to proton instability~\cite{Arnold:2012sd}. 

In general, there are three types of potentially $B$-violating term that can appear in the (renormalisable) potential of an extended scalar sector~\cite{Arnold:2012sd}. Given at least two new scalars $X_1$ and $X_2$, there may be (i) 3-scalar terms of the form $X_1 X_1 X_2$; (ii) 4-scalar terms of the form $X_1 X_1 X_1 X_2$; and (iii)  4-scalar terms of the form $X_1 X_1 X_1 H$ or $X_1 X_1 X_2 H$. In Ref.~\cite{Assad:2017iib} it was shown that {\em any} of the scalar leptoquarks capable of explaining the $B$ anomaly data lead to dangerous tree-level proton decay, either through such terms in the scalar potential or through diquark operators.\footnote{The vector leptoquarks $U_1$ and $U_3$, however, are naturally free of tree-level proton decay. Of the scalar leptoquarks, only the $R_2$ state (which has SM quantum numbers $(\bar{\mathbf{3}},\mathbf{2})_{7/6}$) does not induce tree-level proton decay -- but this leptoquark predicts $R_{K^{(\ast)}} > R_{K^{(\ast)}}^{\text{SM}}$, and so does not provide a good model for the $B$-physics anomalies (on its own). However, all three of these leptoquarks -- $R_2$, $U_1$, and $U_3$ -- admit dimension-5 operators that violate baryon number and facilitate proton decay, schematically of the form $QQH\Phi/\Lambda$, where $\Phi$ denotes the leptoquark. In the case of the $U_1$ leptoquark, this dimension-5 operator is forbidden when the $U_1$ is embedded in, say, a Pati-Salam grand unified theory.  }
In our model there are two new scalars, $S_3$ and $\Phi$. By charging the $S_3$ under $U(1)_X$ one prevents diquark operators as discussed above, and
nor are there any permissible terms in the scalar potential that violate $B$. (Simply by adding up hypercharges, one can rule out operators of all three types listed above.)

Finally, we consider the possibility that baryon number could be violated due to higher-dimension non-renormalisable operators.
Schematically, baryon number could be violated by dimension-5 operators
involving two quarks and two scalars, one of which is the leptoquark. If the other scalar is the Higgs then such an operator cannot possibly be $U(1)_X$ invariant. Neither do there exist invariant operators of the form $QQXX^{(\dagger)}$ for $X\in \{S_3,\Phi\}$.
However, there is potentially a gauge-invariant dimension-5 operator of the form 
\begin{equation} \label{eq:5diquark}
\frac{c_{ij}}{\Lambda}\overline{Q^c_i} (i \sigma_2 \sigma_a S_3^a)^\dagger Q_j \Phi^{(\dagger)},
\end{equation}
where $\Lambda$ denotes the cut-off scale of our BSM effective theory (somewhere beyond the scales associated to the $S_3$, $\Phi$, and $X_\mu$ particles). This operator violates baryon number -- indeed, it reduces precisely to the diquark operator once $\Phi$ acquires its vev.

One might think that this dimension-5 operator could also be forbidden, 
simply by choosing a charge $\hat{Q}_\Phi \notin \{-1,+1\}$.
However, things are not so simple; the charge $\hat{Q}_\Phi$ cannot be freely chosen if our low-energy leptoquark model is to embed in a microscopic theory that can account for neutrino oscillation data, as follows.

As for any model in which $L_\mu - L_\tau$ is gauged, Majorana mass terms are necessary to reproduce the PMNS matrix (see {\em e.g.}~\cite{Heeck:2011wj,Crivellin:2015lwa,Bauer:2018onh}), at least if we assume that a traditional see-saw mechanism is responsible~\cite{minkowski1977mu}. If we include three right-handed neutrinos with charges $0$, $+1$ and $-1$, which allows strictly diagonal Dirac masses via Yukawa couplings to the Higgs, then a SM singlet scalar field $\Phi$ with $U(1)_X$ charge of $\pm 1$ is needed in order to allow off-diagonal Majorana mass terms that involve the first family leptons. The most economical choice is for that scalar field $\Phi$ to be the sole BSM scalar field in the model.\footnote{A richer setup was explored in~\cite{Heeck:2011wj}, in which $U(1)_X$-charged scalar fields transforming as electroweak doublets were also considered in the $U(1)_X$-breaking sector, allowing off-diagonal Dirac mass terms for neutrinos.} In that case there are two zero entries in the 3-by-3 right-handed Majorana neutrino mass matrix at the renormalisable level (in the (2,2) and (3,3) positions), which permits a phenomenologically successful description of neutrino oscillations~\cite{Lavoura:2004tu,Lashin:2007dm}. But this charge assignment $\hat{Q}_\Phi = \pm 1$  immediately allows for the dimension-5 diquark operator (\ref{eq:5diquark}), leading to $B$-violating effects suppressed by one power of the ratio $v_\Phi/\Lambda$. The EFT cutoff scale $\Lambda$ must therefore be pushed high enough to suppress proton decay.

\subsection{The \texorpdfstring{$U(1)_X$}{U(1)X} sector} \label{sec:Zprime}

The role of the SM singlet scalar field $\Phi$ is to spontaneously break $U(1)_X$ by acquiring a non-zero vev, and thus give mass to the corresponding gauge boson, which we denote by the field $X_\mu$ (but we will usually refer to the corresponding particle as a $Z^\prime$ in the text).
From Eq. (\ref{eq:LQ-quark-lepton}), with \(\mu_\Phi, \lambda_\Phi > 0\), we find that \(\braket{\Phi} = v_\Phi/\sqrt{2} = \sqrt{\mu_\Phi^2/ 2 \lambda_\Phi}\).
Writing \(\Phi(x) = (v_\Phi + \varphi(x)) / \sqrt{2}\), one finds that
the $Z^\prime$ boson gets a mass of
\begin{equation}
M_{X} = g_X |\hat{Q}_\Phi| v_\Phi,
\end{equation}
while the scalar field $\varphi$ has a mass
\begin{equation}
M_\Phi^2 = 2 \mu_\Phi^2 = 2 \lambda_\Phi v_\Phi^2
\end{equation}
Note that there is no $Z$-$Z^\prime$ mass-mixing in this model, because the Higgs is uncharged under $U(1)_X$. However, as we will soon see, there is loop-induced kinetic mixing between the $Z$ and $Z^\prime$, which induces couplings of the $Z^\prime$ to quarks.

Of the SM fermions, the $Z^\prime$ boson has tree-level couplings only to leptons, from Eq. (\ref{eq:LQ-quark-lepton}). Explicitly, we have the couplings
\begin{equation}
\mathcal{L}_{\ell Z^\prime} = g_X X_\rho \left( \ \overline{\mu} \gamma^{\rho} \mu - \overline{\tau} \gamma^{\rho} \tau + \overline{\nu}_i \gamma^\rho (U \xi U^\dagger)_{ij} P_L \nu_j \right),
\end{equation}
where all fermion fields are written in the physical mass basis, and $\xi = \text{diag}(0,1,-1)$.

In the lagrangian (\ref{eq:LQ-quark-lepton}) we include a kinetic mixing term $\mathcal{L}\supset \frac{\sin \epsilon}{2} B_{\mu\nu}X^{\mu\nu}$ between the hypercharge and $U(1)_X$ gauge fields. Even if the tree-level kinetic mixing were small there will be radiative contributions of order 
\begin{equation}
\delta \sin\epsilon \sim -\frac{g_Xg'}{16\pi^2} \ln  \frac{\mu^2}{M_{S_3}^2}
\end{equation}
at one-loop, where $\mu$ is the renormalisation scale and $M_{S_3}$ denotes the leptoquark mass.
The physical hypercharge and $Z^\prime$ gauge bosons are the eigenstates of the matrix of kinetic terms, $B_\mu^\prime = B_\mu + \sin \epsilon X_\mu$ and $X_\mu^\prime = \cos \epsilon X_\mu$. Substituting the inverse transformation into the lagrangian and expanding in the physical fields $B_\mu^\prime$ and $X_\mu^\prime$, 
one finds that the couplings of the physical hypercharge boson $B_\mu^\prime$ are unchanged by this kinetic mixing.\footnote{This should be contrasted with the effects of mass-mixing between $B_\mu$ and $X_\mu$, which occurs for models in which the Higgs is charged under $U(1)_X$ (see {\em e.g.}~\cite{Bandyopadhyay:2018cwu,Allanach:2018lvl,Allanach:2019iiy}). In that case, the $Z$ boson acquires corrections to its couplings that may be flavour non-universal.} The $X_\mu^\prime$ couplings acquire a subleading contribution proportional to hypercharge, giving small flavour-universal couplings of the $Z^\prime$ to quarks and thus to protons, proportional to $\tan \epsilon \ll 1$.\footnote{For muon and tau leptons, the kinetic mixing induces only a small correction compared to the direct couplings to the $Z^\prime$, which can be neglected.} At a generic point in parameter space, this would give bounds on $g_X$ and $M_{S_3}$ from direct $Z^\prime$ production at the LHC.
Ultimately, however, we will identify a finitely natural limit in which the $Z^\prime$ decouples completely from the phenomenology, in which these bounds will be unimportant.

%%%%%%%%%%%%%%%%%%%%%%%%%%%%%%%%%%

\section{Finite naturalness of the scalar sector } \label{sec:naturalness}

There are three heavy BSM particles in our theory; the heavy gauge boson $X_\mu$, the SM singlet $\Phi$, and the $S_3$ leptoquark.
These particles can run in loops which induce corrections to the masses of SM particles. In particular, the Higgs boson, being the only scalar in the SM, will receive corrections to its squared mass that grow quadratically with the heavier masses in our BSM theory.
These corrections are calculable within our model, if one assumes that there are no states to be discovered at yet higher energies (or neglects their contributions). Moreover, the additional scalars $\Phi$ and $S_3$ will themselves receive calculable mass corrections, and we might expect the lighter of the two to be dragged up by such a loop correction to the higher mass scale. 
A theory in which the finite loop corrections do not need to be fine-tuned to produce the observed values of couplings and masses is `finitely natural'~\cite{Farina:2013mla}.

To be concrete, if we denote the one-loop finite correction to the Higgs mass-squared by $\delta M_H^2$, then finite naturalness requires that
\begin{equation}
\delta M_H^2 \lesssim M_H^2 \times \Delta, \label{eq:finite nat H}
\end{equation}
where $\Delta$ indicates the degree of fine-tuning that one is willing to tolerate. For the SM itself, Eq. (\ref{eq:finite nat H}) is satisfied with $\Delta = 0.13$ at a renormalisation scale $\mu = m_t$~\cite{Farina:2013mla}, meaning that the SM is a finitely natural theory.
One might view this lack of tuning as merely a coincidence, or, more optimistically, as an attractive feature of the SM that we should seek to preserve when we go beyond the SM.

Because the $b\to s \ell \ell$ anomalies occur in rare processes that are loop and GIM suppressed in the SM, the scale of new physics required is rather heavy, corresponding to a mass over coupling of order $30-40$ TeV or so. One might therefore expect the finite corrections to the Higgs mass to be large, and require significant fine-tuning for the observed mass of $M_H \approx 125$ GeV to be finitely natural. We here confront this expectation,
focussing on a r\'egime in which there is a strong hierarchy between the three mass scales, 
\begin{equation}
M_H \ll M_{S_3} \ll M_\Phi . \label{eq:mass hierarchy}
\end{equation}
We will show that this mass hierarchy is in fact finitely natural for certain regions of parameter space that can accommodate the $B$ anomalies (at the best fit point). In particular, in \S \ref{sec:decoupling} we consider the limit where 
\begin{equation} \label{eq:decoupling}
g_X \to 0 \; \text{ and } \; v_\Phi \to \infty, \; \text{ such that } M_X \text{ remains finite},
\end{equation}
where recall $M_X = g_X |\hat{Q}_\Phi| v_\Phi $. In this limit the low-energy phenomenology is much simplified, as we discuss in \S \ref{sec:phenomenology}, because the $Z^\prime$ boson decouples.

In this decoupling limit, the Higgs mass (and the hierarchy (\ref{eq:mass hierarchy})) can be finitely natural provided the leptoquark is not too heavy (but heavy enough to evade the LHC direct search bounds), and provided its coupling $\alpha_3$ to the top quark is not too large with respect to $\alpha_2$. 
These results therefore have interesting implications for model-building frameworks for the $B$ anomalies which invoke a $U(2)$-like flavour structure, for which couplings of new physics to the third family are very large. The implication is that, for such a flavour structure, the leptoquark should not be much heavier than 5 TeV if we do not wish to destabilize the Higgs mass, which is not too far from the current reach of LHC direct searches. We will discuss the implications of finite naturalness for more general classes of explanations of the $B$ anomalies in a follow up paper.

%%%%%%%%%%%%%%%%%%%%%%%%%%%%%%%
\subsection{Higgs mass stability}  \label{sec:Hmass}

\begin{figure}
\includegraphics[width=\textwidth]{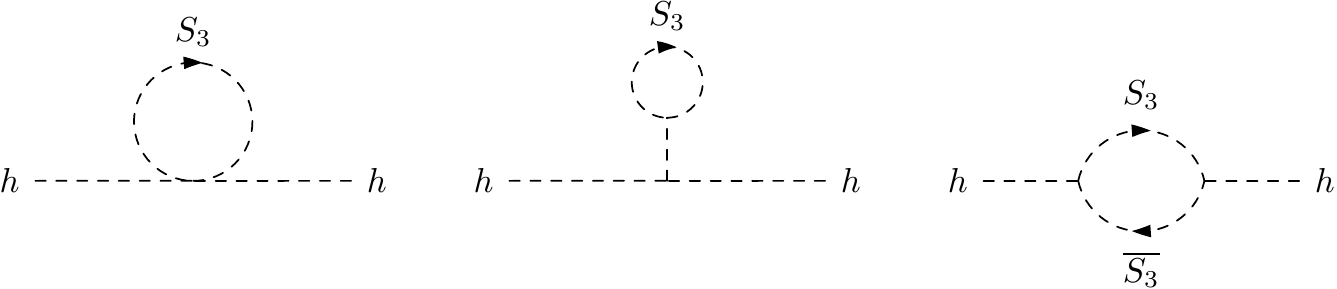}
\caption{One-loop Feynman diagrams contributing to the Higgs mass, due to the $S_3$ leptoquark running in the loop.}
\label{fig:DeltaMH}
\end{figure}

There are one-loop corrections to the Higgs mass due to the $S_3$ leptoquark running in a loop (see Fig.~\ref{fig:DeltaMH}).\footnote{We make the simplifying assumption throughout that $\lambda_{\Phi H}=0$. This assumption is arguably natural because $\lambda_{\Phi H}$ is not generated at one-loop. Moreover, it is pragmatic for two reasons.
Firstly, if $\lambda_{\Phi H}\neq 0$ then the symmetry breaking pattern is potentially more complicated, with the vevs of $H$ and $\Phi$ being determined by the true minimum of the scalar potential. Secondly, $\lambda_{\Phi H}\neq 0$ would give a potentially large tree-level correction to the Higgs mass. }
In the limit that $M_{S_3} \gg M_H$, we find (see Appendix \ref{app:mass_corrections} for details of the computation)
\begin{equation} \label{eq:higgs_mass_correction}
(\delta M_H^2)_{HS} = -\frac{9 M_{S_3}^2}{16\pi^2} \left[ \left(\lambdaHS + \lambdaHSprime\right) \left(1 + \ln \frac{\mu^2}{M_{S_3}^2}\right) + \BigO{\frac{v^2}{M_{S_3}^2}} \right] \,,
\end{equation}
where $\mu$ is the renormalisation scale.
It appears that this one-loop correction can be made arbitrarily weak by taking the tree-level couplings  $\lambda^{(\prime)}_{HS}$ to be small.
However, the coupling $\lambda^{\prime}_{HS}$ is generated at one-loop
due to the Yukawa-like couplings $\alpha_i$,\footnote{It turns out that to leading order only $\lambda^\prime_{HS}$ is generated by one-loop diagrams; \(\lambdaHS\) is suppressed relative to it by \(m_b^2 / m_t^2 \sim 10^{-3}\).} whose sizes are fixed by the $b\to s \ell\ell$ anomalies.

We estimate the size of a radiatively-generated $\lambda^\prime_{HS}$ by computing its one-loop beta function (see Appendix \ref{app:loop_induced}), which may be substituted into (\ref{eq:higgs_mass_correction}) to give a lower estimate of the correction (\ref{eq:higgs_mass_correction}),
\begin{equation} \label{eq:MHcorrection1}
(\delta M_H^2)_\text{anomaly} \approx \frac{9 G_F M_{S_3}^2 m_t^2 |\sum_q \alpha_q V_{tq}|^2}{32\sqrt{2} \pi^4}  \left(1 + \ln \frac{\mu^2}{M_{S_3}^2}\right)  \ln \frac{\mu^2}{m_t^2} + \dots,
\end{equation}
where $M_t$ is the top mass, and the subscript `anomaly' now indicates that this estimated loop correction is tied to the size of the $b\to s \ell\ell$ anomalies.
The finite naturalness criterion (\ref{eq:finite nat H}) then implies
\begin{equation} \label{eq:HNat1}
M_{S_3} \lesssim 
\begin{cases}
	\frac{4.7 \text{~TeV}}{|\alpha_3 + V_{ts} \alpha_2|\sqrt{\ln M_{S_3}^2 / m_t^2}} \sqrt{\Delta} &\text{ for } \mu = M_{S_3}, \\
	\frac{540 \text{~GeV}}{|\alpha_3 + V_{ts} \alpha_2| \sqrt{74 + \ln M_{S_3}^2 / m_t^2}} \sqrt{\Delta} &\text{ for } \mu =10^{16} M_{S_3} \sim M_\text{Planck},
\end{cases}
\end{equation}
where in each case $\mu$ indicates the renormalisation scale used.\footnote{The rationale in presenting two bounds is to estimate how strongly the finite naturalness corrections depend on the scale; the first bound would be appropriate if relevant new physics were to enter just beyond the scale of the leptoquark, while the second bound would be appropriate if there were a `desert' of no new physics all the way to the Planck scale.}
Note that this finite naturalness bound depends on the scale of new physics required to explain the $B$-anomalies, through both the couplings $\alpha_i$ and the leptoquark mass.

The calculation (\ref{eq:MHcorrection1}) is the combination of two one-loop integrals, and so amounts to a two-loop correction to the Higgs mass. We should therefore consider other corrections at two-loop order. Totally generically, there is a two-loop Higgs mass correction due to electroweak gauge bosons running in the loop, for any BSM theory with an electroweakly-charged scalar~\cite{Farina:2013mla}. For the $S_3$ leptoquark, that correction is
\begin{equation} \label{eq:higgs_2loop}
(\delta M_H^2)_\text{EW} = - \frac{\alpha^2 M_{S_3}^2}{32\pi^2} \left( \frac{18}{s_W^4} + \frac{1}{c_W^4} \right) \left( 3 \ln^2 \frac{M_{S_3}^2}{\mu^2} + 4 \ln \frac{M_{S_3}^2}{\mu^2} + 7 \right),
\end{equation}
where $\alpha$ is the fine structure constant. For this correction, the finite-naturalness criterion (\ref{eq:finite nat H}) implies
\begin{equation}  \label{eq:HNat2}
M_{S_3} \lesssim 
\begin{cases}
	5.8 \text{~TeV} \sqrt{\Delta} &\text{ for~ } \mu = M_{S_3}, \\
	120 \text{~GeV} \sqrt{\Delta} &\text{ for~ } \mu = 10^{16} M_{S_3} \sim M_\text{Planck},
\end{cases}
\end{equation}
for the same pair of renormalisation scales used above.
We stress that this second Higgs mass correction is inevitable in {\em any} model featuring the $S_3$ leptoquark, simply due to its quantum numbers under the electroweak gauge group. This bound, unlike (\ref{eq:HNat1}), is independent of the Yukawa-like couplings $\alpha_i$, depending only on the leptoquark mass.

%%%%%%%%%%

There is also a two-loop correction to the Higgs mass that is very specific to our model, featuring the heavy $X$ boson being exchanged within a muon or tauon loop. However, these corrections scale like $\delta M_H^2 \sim \frac{m_{\mu,\tau}^2}{v^2}\frac{g_X^2 M_X^2}{(4\pi^2)^2}$, and so they vanish in the `decoupling limit' specified by Eq.~(\ref{eq:decoupling}).

%%%%%%%%%%

Typically, one of the two bounds (\ref{eq:HNat1}) or (\ref{eq:HNat2}) will be dominant in a given region of parameter space, for a given choice of $\Delta$.\footnote{A refined estimate of the finite naturalness bound would be obtained by doing a full calculation of the two-loop Higgs mass correction. Here, we are content to provide only a rough estimate of the important two-loop Higgs mass contributions for the parameter space regions of interest.
}
 In Fig.~\ref{fig:MLQ}, we plot the stronger of the two finite naturalness bounds at any given point in the parameter space. We see that for $\alpha_2 / \alpha_3 \ll 1$ and larger leptoquark masses, the bound (\ref{eq:HNat1}) tends to dominate, while for light leptoquarks and/or a larger $\alpha_2$ contribution to the $B$-anomalies, the purely electroweak correction (\ref{eq:HNat2}) can dominate.

One should also study the stability of the leptoquark mass, which will receive finite loop corrections that are sensitive to the mass scale of the heavy states associated with the $U(1)_X$ gauge sector. We find that, in the decoupling limit (\ref{eq:decoupling}), finite naturalness of the leptoquark mass gives only very weak constraints on the parameters of the model, that have no impact on the phenomenology. We therefore relegate these calculations to Appendix~\ref{app:LQ2}.

\subsection{Decoupling the $U(1)_X$ sector} \label{sec:decoupling}

In the limit given by Eq. (\ref{eq:decoupling}), in which we take $g_X\to 0$\footnote{The reader might worry about taking $g_X \to 0$, because there is growing evidence for the weak gravity conjecture which suggests that one cannot in fact take a gauge coupling to be arbitrarily small~\cite{ArkaniHamed:2006dz}. However, the lower limit on $g_X$ is expected to be of order $m/M_\text{Planck}$ for order-1 charges, where $m$ is the mass of a $U(1)_X$ charged particle in the spectrum. Thus, for all practical purposes we can take $g_X$ as small as we need without running the risk of violating this bound from weak gravity. } and $v_\Phi\to \infty$ while keeping their product (and thus $M_X$) finite, all the finite loop corrections to either $M_H^2$ or $M_{S_3}^2$ that involve either the $X$ boson or the heavy scalar $\Phi$ running in the loop tend to zero (at least as fast as $g_X^2 M_X^2$).
Moreover, even though the $Z^\prime$ mass remains finite, all the physics effects of exchanging the $Z^\prime$ depend only on the parameter $v_\Phi$ (as long as the $Z^\prime$ is heavy enough not to be produced on resonance), and so disappear in the limit $v_\Phi \to \infty$, regardless of the coupling. (Note that the mass of the scalar field $M_\Phi = \sqrt{2\lambda_\Phi} v_\Phi \to \infty$ in this limit.)
Thus, there is a limit in which the $X$ and $\Phi$ are heavy enough that they totally decouple from the low-energy phenomenology, without inducing large finite mass corrections to the leptoquark or the Higgs.

Indeed, this is not all that surprising from the point of view of symmetry. If one takes to zero the various parameters that mix the $U(1)_X$ sector with the SM + $S_3$ sector, one finds that the lagrangian (\ref{eq:LQ-quark-lepton}) exhibits an enhanced $G=(\text{Poincar\'e})^2$ spacetime symmetry, which acts by performing independent Poincar\'e transformations within each sector. Thus, by a famous argument of 't Hooft~\cite{tHooft:1979rat}, one expects this limit to be radiatively stable, {\em i.e.} one expects the various mass corrections to be small (and indeed to vanish in the limit of complete decoupling).\footnote{Note also that in this limit of decoupling the $Z^\prime$ the radiatively-induced kinetic mixing between the $Z^\prime$ and hypercharge gauge bosons also goes to zero, which can again be traced back to the radiative stability of decoupling the $U(1)_X$ sector from the SM + $S_3$ sector. } 

However, we have seen that the finite corrections (\ref{eq:MHcorrection1}) to the Higgs mass arising from leptoquark loops cannot be made arbitrarily small, because they are tied to the flavour anomalies through the dependence on the couplings $\alpha_{2,3}$. There is also an unavoidable two-loop Higgs mass correction (\ref{eq:higgs_2loop}) due to electroweak gauge bosons running in the loop, which have fixed couplings to the $S_3$. Both these Higgs mass contributions arise generically for an $S_3$ leptoquark that can explain the $b\to s \ell \ell$ anomalies. We visualize the `bounds' from finite naturalness of the Higgs mass, which translate into an upper bounds on $M_{S_3}$, in Figs.~\ref{fig:a3a2} and~\ref{fig:MLQ}. For suitably small values of the fine-tuning parameters $\Delta$, the scalar mass hierarchy (\ref{eq:mass hierarchy}) is finitely natural.

\section{Phenomenological analysis in the decoupling limit} \label{sec:phenomenology}

We now turn to the phenomenology of our charged leptoquark model. At a general point in the parameter space the phenomenology is rich, with constraints associated both to the leptoquark and $Z^\prime$ dynamics. The latter includes constraints from direct $Z^\prime$ production at the LHC, neutrino trident production, and four-lepton production. However, in the previous section we identified a limit in which the $Z^\prime$ decouples from the phenomenology. In this section we analyse the phenomenology in this ``decoupling limit'' only. 

The parameter space is reduced from the set of 14 parameters in (\ref{eq:parameters}) to the smaller 5-parameter set $\{ \alpha_i, M_{S_3}, \lambda_S\}$. Furthermore, for the heavy leptoquark masses of interest,\footnote{The leptoquark mass will be at least 1.7 TeV or so to avoid direct search bounds - see \S \ref{sec:LHCdirect}.} the leptoquark quartic self-coupling $\lambda_S$ plays little role in the phenomenology, and so we neglect it henceforth.
We are thus left with the following set of four parameters,
\begin{equation} \label{eq:reduced parameters}
\{ \alpha_1, \alpha_2, \alpha_3, M_{S_3} \},
\end{equation}
which determine the leptoquark phenomenology.\footnote{Note that while most of the parameters we have removed are taken to zero (or infinity) in our decoupling limit, the coupling $\lambda'_{HS}$ remains finite, but may be written in terms of $\alpha_3$ using the one-loop estimate of Appendix \ref{app:loop_induced}.}

The main experimental constraints on our model in this decoupling limit will be threefold, coming from (i) the fit to the neutral current $B$ anomaly data (\S\ref{sec:B anomalies}), (ii) neutral meson mixing (\S \ref{sec:B mixing}), and (iii) direct leptoquark searches at the LHC (\S \ref{sec:LHCdirect}). 
We visualize this combination of constraints in two ways in Figs.~\ref{fig:a3a2} and~\ref{fig:MLQ}; in the former, we fix benchmark values of the leptoquark mass and plot the constraints in the $\alpha_2$ {\em vs.} $\alpha_3$ plane.
Then, to study the mass-dependence, in Fig.~\ref{fig:MLQ} we fix the value of $\Delta C^\mu_9 = -\Delta C^\mu_{10}$ using the central value of the global fit~\cite{Alguero:2019ptt} to the $B$ anomaly data, and plot the remaining constraints in the $\alpha_2/\alpha_3$ {\em vs.} $M_{S_3}$ plane. In both plots we assume for simplicity that the $\alpha_i$ couplings are all real. We also include the `bounds' (\ref{eq:HNat1}) or (\ref{eq:HNat2})  from finite naturalness of the Higgs mass, calculated with $\mu=M_{S_3}$, which we see have interesting implications on the favoured parameter space.

\subsection{Neutral current $B$ anomalies} \label{sec:B anomalies}

Assuming that the new physics behind the collection of experimental discrepancies in $b \to s \ell\ell$ occurs only in the muonic channel, we consider the following terms in the WET,
\begin{equation}
\mathcal{L}_{bs\mu\mu}^\text{NP} = \frac{4G_F}{\sqrt{2}} V_{tb} V_{ts}^\ast \left( \Delta C_9^\mu \mathcal{O}_9^\mu + \Delta C_{10}^\mu \mathcal{O}_{10}^\mu \right),
\end{equation}
where
\begin{align}
\mathcal{O}_9^\mu &= \frac{\alpha}{4\pi} \left( \overline{s}_L \gamma_\mu b_L \right) \left(\overline{\mu} \gamma^\mu \mu \right), \\
\mathcal{O}_{10}^\mu &= \frac{\alpha}{4\pi} \left( \overline{s}_L \gamma_\mu b_L \right) \left(\overline{\mu} \gamma^\mu \gamma_5 \mu \right).
\end{align}
Integrating out the $S_3$ leptoquark and matching onto the WET, we obtain
\begin{equation} \label{eq:DeltaC9}
\Delta C^\mu_9 = -\Delta C^\mu_{10} = \frac{\pi}{\sqrt{2} G_F M^2_{S_3} \alpha_\text{EM}} \left( \frac{\alpha_3 \alpha_2^*}{V_{tb} V^\ast_{ts}} \right) \,.
\end{equation}
As anticipated (and intended), the leptoquark generates only the operator with left-handed quark current and left-handed lepton current, and so our model predicts \(R_K = R_{K^*}\), which gives a good fit to the $B$-anomaly data~\cite{Aebischer:2019mlg,Alguero:2019ptt}.

\begin{figure}
\includegraphics[width=0.95 \textwidth]{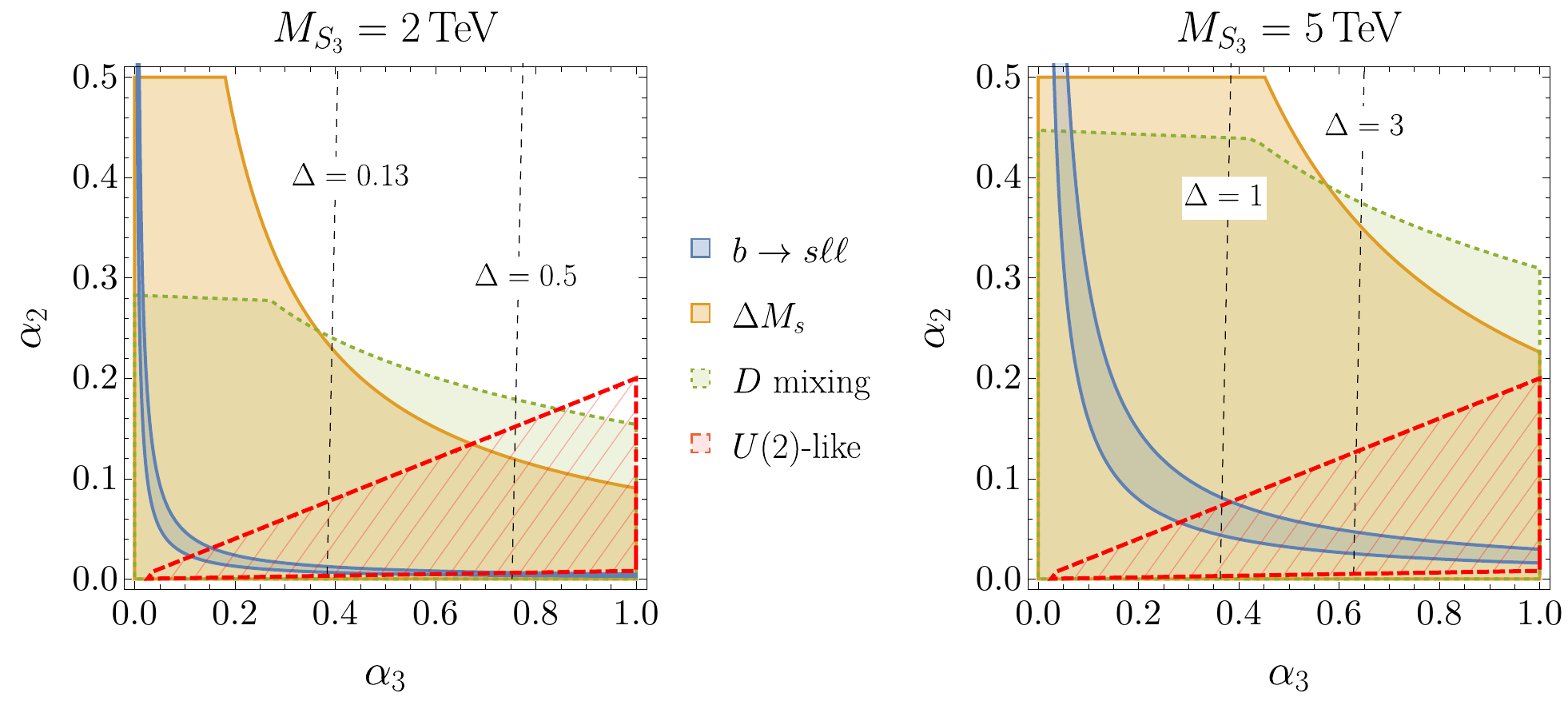}
\caption{ Constraints on the leptoquark couplings to $s \mu$ ($\alpha_2$) and $b \mu$ ($\alpha_3$), for fixed leptoquark masses of 2 TeV (left) and 5 TeV (right), both of which are above the mass range excluded by LHC direct searches. In addition to the 2$\sigma$ contours from a global fit to the $b\to s \ell\ell$ data (blue), from $B_s$ mixing constraints (orange), and from $D$ mixing constraints (green, dotted), we plot contours from finite naturalness of the Higgs mass, in this case coming from Eq. (\ref{eq:HNat1}). Finally, the red hatched regions are compatible with a $U(2)$-like flavour structure, for which \(\alpha_2 / \alpha_3 \in [0.2,5] \V{ts}\). From the left plot, we see that a lighter mass leptoquark requires significantly less Higgs-mass tuning; the contour with $\Delta=0.13$ indicates comparable tuning to that in the SM. For the 5 TeV leptoquark, we see that a $U(2)$-like structure can only accommodate the 2$\sigma$ fit to the $B$-anomalies with $\Delta \gtrsim 1$, meaning the the finite loop correction to the Higgs mass is greater than 125 GeV.
\label{fig:a3a2}}
\end{figure}

A recent fit for a purely \(\Delta C_9^\mu = - \Delta C_{10}^\mu\) scenario \cite{Alguero:2019ptt} (specifically, we use the results of the 2020 addendum, which includes the updated \(P'_5\) result \cite{Aaij:2020nrf}) finds \( \Delta C_9^\mu=-\Delta C_{10}^\mu=-0.50\) as the best-fit central value, and \(-0.69 \leq (\Delta C_9^\mu=-\Delta C_{10}^\mu) \leq -0.32\) as the $2\sigma$ range.\footnote{Various other fitting methodologies (see \cite{Kowalska:2019ley,Ciuchini:2019usw,Arnan:2019uhr,DAmico:2017mtc,Alok:2019ufo}) yield similar results.}
Assuming that $\alpha_3 \alpha_2^*$ is real and positive, this implies that
\begin{equation}
\frac{M_{S_3}}{\sqrt{\alpha_3 \alpha_2^*}} = 34.7 \text{~TeV}
\end{equation}
as a central value, or 
\begin{equation}
\label{eq:NCanomaly_2sig}
30.4 \text{~TeV} \leq \frac{M_{S_3}}{\sqrt{\alpha_3 \alpha_2^*}} \leq 44.6 \text{~TeV}
\end{equation}
as the corresponding $2\sigma$ interval.
We have used \texttt{CKMfitter} (Summer 2019 version) \cite{Charles:2004jd,CKMfitter:Summer19} for the CKM input, and neglected the small imaginary part of $V_{ts}$. For the fixed values $M_{S_3}=2,\; 5\text{~TeV}$, this 2$\sigma$ band is plotted (blue) in Fig.~\ref{fig:a3a2}.

\subsection{Neutral meson mixing} \label{sec:B mixing}

As for any model that fits the $b \to s \ell\ell$ anomalies, an important constraint comes from $B$ meson mixing amplitudes, in particular from the $B_s$ meson mixing observable \(\Delta M_s\) \cite{DiLuzio:2017fdq,DiLuzio:2019jyq}.
In our model, as a consequence of gauging the leptoquark under $L_\mu - L_\tau$, we have forced a direct connection between $B_s$ mixing and the flavour anomalies, both of which are governed by the combination of couplings $\alpha_3 \alpha_2^*$. (For generic leptoquark couplings to leptons, one finds a sum over all lepton generations in the box diagram that contributes to the mixing, thus breaking this direct connection.)

\begin{figure}
\includegraphics[width=0.65 \textwidth]{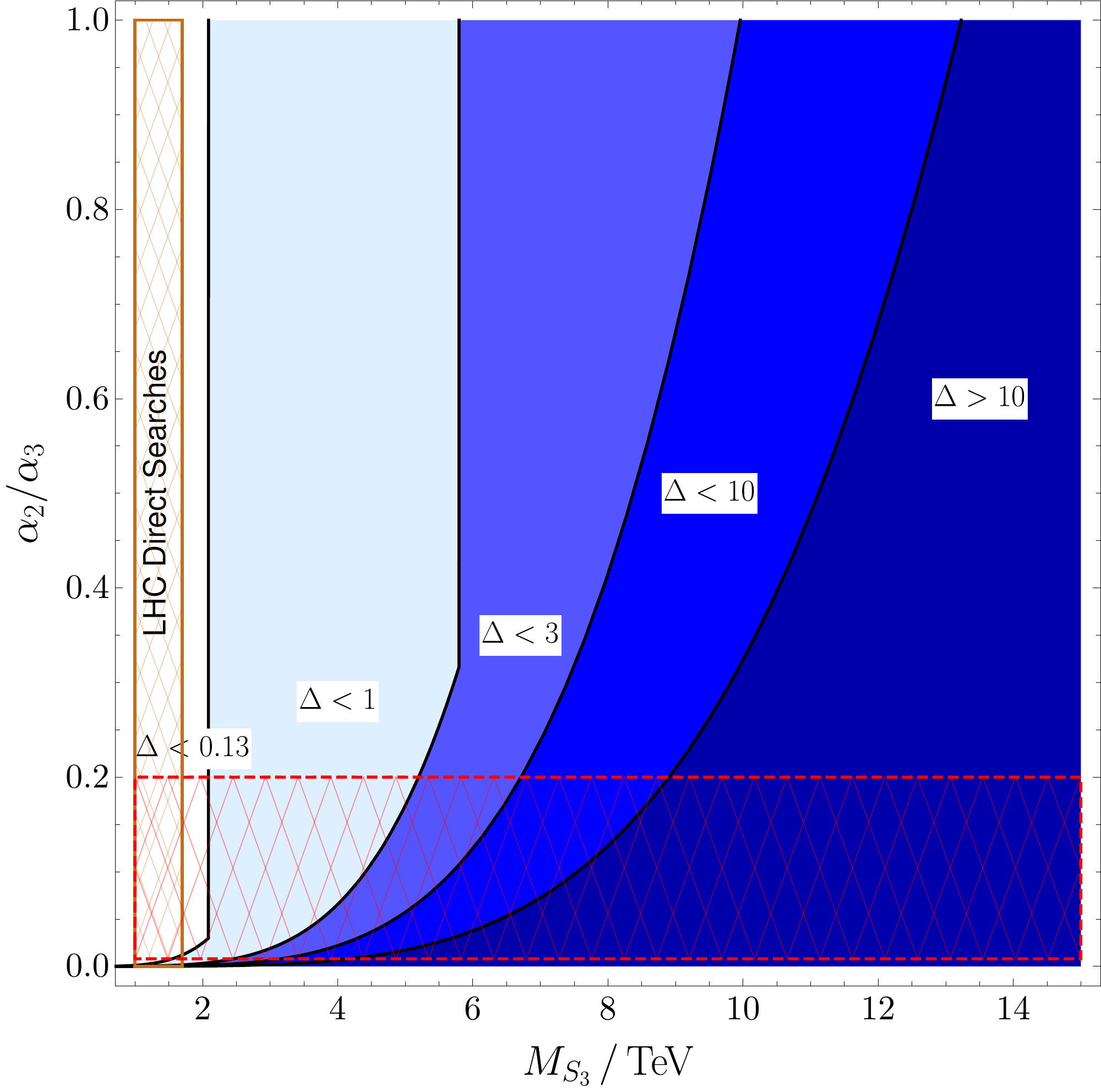}
\caption{Constraints on the ratio $\alpha_2/\alpha_3$ {\em vs.} the leptoquark mass, with $\Delta C^\mu_9 = -\Delta C^\mu_{10}$ everywhere fixed to the central value of the global fit~\cite{Alguero:2019ptt} to $B$ anomaly data. The brown hatched region ($M_{S_3} \leq 1.7$ TeV) is excluded by ATLAS direct searches~\cite{Aad:2020iuy}.
The constraint from $B_s$ meson mixing is not visible on the plot, since at the best fit point for the anomalies it only imposes an upper bound of \(M_{S_3} \leq \SI{55}{\TeV}\). The blue shaded regions indicate the degree of fine-tuning of the finite Higgs mass corrections, with $\Delta > 1$ indicating that the finite loop corrections exceed 125 GeV. The region with $\Delta \leq 0.13$ exhibits a similar degree of naturalness to the SM, for which only a sliver of parameter space remains compatible with direct searches.
Finally, the red hatched region shows the \(U(2)\)-like region where \(\alpha_2 / \alpha_3 \in [0.2,5] \V{ts}\), which requires a tuning $\Delta >1$ for $M_{S_3}>5.2$ TeV.
\label{fig:MLQ}}
\end{figure}

Using the most recent SM predictions for \(\Delta M_s\) \cite{DiLuzio:2019jyq}, which is
\begin{equation}
\Delta M_s^\text{SM} = \left(18.4^{+0.7}_{-1.2}\right)\,\text{ps}^{-1},
\end{equation}
we obtain the following constraint on the parameters of our model,
\begin{equation}
0.88 \leq \left| 1 + \left(\frac{\alpha_3 \alpha_2^*}{M_{S_3} / 6.3 \text{~TeV}}\right)^2 \right| \leq 1.08
\end{equation}
at 2$\sigma$. Assuming that \(\alpha_3 \alpha_2^*\) is real and positive, this reduces to the bound
\begin{equation} \label{eq:Bs_bound}
\frac{M_{S_3}}{\alpha_3 \alpha_2^*} \geq 22\text{~TeV}.
\end{equation}
For the fixed values $M_{S_3}=2,\; 5\text{~TeV}$, this bound is plotted (orange) in Fig.~\ref{fig:a3a2}, and is consistent with the whole region plotted in Fig.~\ref{fig:MLQ}.

There are similar bounds arising from other neutral meson mixing amplitudes, which constrain different combinations of the leptoquark couplings $\alpha_i$, depending on the quark content of the meson. Constraints from $B_d$ \cite{DiLuzio:2019jyq} and kaon\footnote{\label{foot:UTfit}For kaon mixing, we use the results from Ref.~\cite{Bona:2007vi} as updated at `La Thuile 2018' by L. Silvestrini, for which slides are available at \url{https://agenda.infn.it/event/14377/contributions/24434/attachments/17481/19830/silvestriniLaThuile.pdf}.  } mixing translate to the bounds
\begin{equation}
\frac{M_{S_3}}{\alpha_3 \alpha_1^*} \geq 91 \text{~TeV} \qquad \text{and} \qquad \frac{M_{S_3}}{\alpha_2 \alpha_1^*} \geq 72 \text{~TeV}
\end{equation}
respectively,
again assuming that the corresponding products $\alpha_i \alpha_j^*$ are real and positive.
If we are in a region of parameter space where the neutral current anomaly is explained (\ref{eq:NCanomaly_2sig}), these two observables can place an upper limit \(|\alpha_1| \leq 0.55\).
In general however, since  $\alpha_1$ is not connected to the flavour anomalies, it can be taken much smaller, making both these bounds from $B_d$ and kaon mixing essentially unimportant.

More importantly, there are contributions to \(D\) meson mixing which cannot be avoided by taking \(\alpha_1\) to be small, since we chose a basis in which the CKM mixing is in the up sector, and so the up quark couples to \(S_3\) through \(\sum_k \alpha_k \V*{1k}\).
The BSM contribution to the effective \(\Delta C = 2\) Hamiltonian is
\begin{equation}
C_D^1 = \frac{5}{128\pi^2} \frac{1}{M_{S_3}^2} \Bigg( \sum_j \alpha_j \V*{2j} \Bigg)^2 \Bigg( \sum_k \alpha_k^* \V{1k} \Bigg)^2 \,.
\end{equation}
We take our bound from the recent \texttt{UTfit} update (see footnote \ref{foot:UTfit}).
For the limiting value \(\alpha_1 = 0\), the \(D\) mixing bound translates to
\begin{equation}
\frac{M_{S_3}}{|(\alpha_2 (\alpha_2 + 0.04 \alpha_3)|} \gtrsim \SI{25}{\TeV} \,, \quad \frac{M_{S_3}}{\sqrt{\alpha_3 \alpha_2} |\alpha_2 + 0.05 \alpha_3|} \gtrsim \SI{25}{\TeV},
\end{equation}
for the real and imaginary parts of \(C_D^1\) respectively.
However, we remark that  non-zero \(\alpha_1\) can lead to cancellations that weaken this constraint, particularly in the $U(2)$-like region with \(\alpha_1 / \alpha_2 \sim \V{td} / \V{ts}\).
It is also the case that there is significant uncertainty in the value of the SM prediction for \(D\) mixing and how much is contributed from short and long distance effects, and so the green shaded regions in Fig.~\ref{fig:a3a2} should be considered subject to these caveats.

To obtain the $B_s$ mixing bound (\ref{eq:Bs_bound}) we assumed that the leptoquark coupling product \(\alpha_3 \alpha_2^*\) was real. Of course, these couplings are in general complex numbers, and any relative phases between the components $\alpha_i$ are constrained by $CP$-violating observables, such as the $CP$ asymmetry $A^\text{mix}_\text{CP}$ \cite{DiLuzio:2017fdq}.
For a leptoquark mass of 5 TeV or lower, for which a $U(2)$-like flavour structure remains finitely natural (see Fig.~\ref{fig:ROFV} and \S \ref{sec:other}), the experimental constraint from $A^\text{mix}_\text{CP}$ is very weak, and the relative phase between $\alpha_2$ and $\alpha_3$ is essentially unconstrained. (The relative phases with respect to $\alpha_1$ are, however, more strongly constrained, by both phases in $B_d$ meson mixing and also by decays sensitive to the imaginary parts of Wilson coefficients, {\em e.g.} $K_L \to \pi^0 \mu^+ \mu^-$. But, since $\alpha_1$ is essentially a free parameter that plays no role in mediating the $B$ anomalies, we shall not consider these phases further in this paper.)

\subsection{Leptoquark pair production at the LHC} \label{sec:LHCdirect}

The third important constraint on our reduced parameter space (\ref{eq:reduced parameters}) comes from direct leptoquark production at the LHC.

The particularities of our flavour structure, which arise from charging the leptoquark under $L_\mu - L_\tau$, make its collider phenomenology rather specific, and a little different from that discussed for other $S_3$ leptoquark solutions to $B$ anomaly data; in particular, decays to third-family leptons are often considered as a dominant channel (see {\em e.g.}~\cite{Faroughy:2016osc,Dorsner:2017ufx}), whereas our charged leptoquark decays only to second-family leptons. The most up-to-date constraints in muonic decay channels come from searches by ATLAS~\cite{Aad:2020iuy} for pair-produced scalar leptoquarks decaying into two muons plus two or more jets, which we use here.

The cleanest (and tightest) constraints come from pair production of the $S_3^{4/3}$ component, each of which decays exclusively to $\overline{d_i} \mu^+$, for some down-type quark $d_i$. (The other components of the $S_3$ have tree-level decays to neutrinos, which generally result in weaker limits.) The specific branching ratios for each final state are determined by the direction of the vector $\alpha_i$, as
\be
\text{BR}(S_3^{4/3} \to \overline{d_i} \mu^+) \approx \frac{|\alpha_i|^2}{\sum_j |\alpha_j|^2},
\ee
in the limit that all down-type quarks and the muon are massless.
In Ref.~\cite{Aad:2020iuy}, limits on the $S_3$ mass are computed for each channel, in each case assuming that the leptoquark decays exclusively into one specific combination of quark flavour and lepton flavour. For our model this is not generically the case, but in the region of parameter space where $\alpha_3 \gg \alpha_{1,2}$, the $S_3^{4/3}$ component decays exclusively to $\overline{b} \mu^+$, and the ATLAS limit is $M_{S_3^{4/3}} \gtrsim \SI{1.7}{\TeV}$~\cite{Aad:2020iuy}.
Conversely, if $\alpha_3 \ll \alpha_{1,2}$, the $S_3^{4/3}$ component decays to $\overline{q}\mu^+$, where $q$ is a light quark ($\overline{d}$ or $\overline{s}$), for which the limit is again $M_{S_3^{4/3}} \gtrsim \SI{1.7}{\TeV}$~\cite{Aad:2020iuy}.\footnote{This bound assumes that $\alpha_1$ is not so large that leptoquark pair production is no longer dominated by gluon fusion, which anyway seems ruled out by the combined meson mixing and neutral current anomaly bound on \(\alpha_1\).}
Since the mass exclusion is roughly the same in both these extreme cases, it seems reasonable to interpolate that the $2\sigma$ bound on the leptoquark mass for some generic value of couplings $\alpha_i$ should be around
\be
M_{S_3^{4/3}} \gtrsim \SI{1.7}{\TeV}.
\ee
This is the direct search bound that we plot in Fig.~\ref{fig:MLQ}.

\subsection{Other constraints} \label{sec:other}

\begin{figure}
\includegraphics[width=0.95\linewidth]{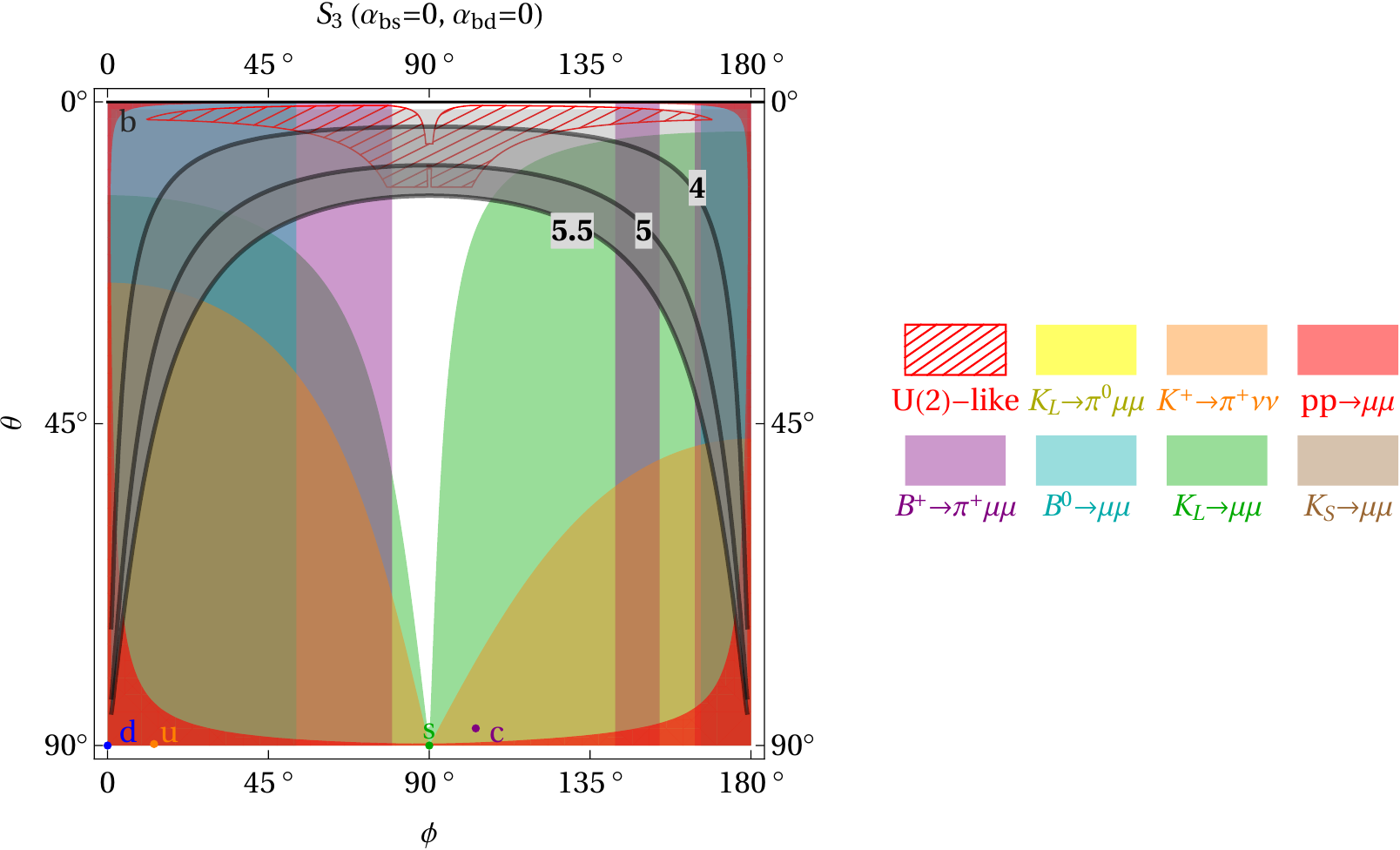}
\caption{ Constraints on the polar angles $\theta$ and $\phi$, which specify the direction of the 3-vector $\alpha_i$ (assumed real) of leptoquark couplings, from a variety of correlated flavour observables (updated from \cite{Gherardi:2019zil} with the latest \(b\to s\ell\ell\) fit). The white region is allowed at 2$\sigma$. The black lines are finite naturalness contours with \(\Delta = 1\), for given values of the leptoquark mass below 5.8 TeV (for which the anomaly-tied correction (\ref{eq:MHcorrection1}) is the dominant one). Above each line, the leptoquark induces a Higgs mass correction greater than 125 GeV or so, and so the model requires fine-tuning. We find that a $U(2)$-like flavour structure is compatible with finite naturalness (\emph{i.e.} $\Delta \lesssim 1$) for $M_{S_3} \lesssim 5$ TeV or so.
When $M_{S_3} \geq 5.8$ TeV there is an inevitable tuning of $\Delta \geq 1$ coming from the purely electroweak 2-loop correction (\ref{eq:higgs_2loop}), which is independent of $(\theta,\phi)$ ({\em i.e.} independent of the flavour structure).
\label{fig:ROFV}}
\end{figure}

There are other constraints on the direction $(\alpha_1,\alpha_2,\alpha_3)$ of flavour violation in quark space, coming from precision measurements of correlated flavour observables. Since our charged leptoquark setup exhibits rank-one flavour violation, with couplings only to muons, the results from Ref.~\cite{Gherardi:2019zil} are directly transferable. In Fig.~\ref{fig:ROFV} we reproduce the constraints from Ref.~\cite{Gherardi:2019zil}, relevant to the particular case of the $S_3$ leptoquark couplings, subject to the simplifying assumption that $\alpha_i \in \mathbb{R}$. The constraints are plotted in the space of polar angles $\theta$ and $\phi$ that define the direction of the $\alpha_i$ vector, {\em viz.} $\alpha_i = \sqrt{\sum_i \alpha_i^2}\; \hat{n}_i$, where $\hat{n}_i^T = (\sin\theta\cos\phi, \sin\theta \sin\phi, \cos\theta)$.

Fig.~\ref{fig:ROFV} includes constraints, as calculated in Ref.~\cite{Gherardi:2019zil}, principally from observables probing $d_i \to d_j \mu^+ \mu^-$ transitions, specifically from $\text{BR}(B_d^0 \to \mu^+ \mu^-)$, $\text{BR}(B^+ \to \pi^+ \mu^+ \mu^-)$, $\text{BR}(K_S \to \mu^+ \mu^-)$, and $\text{BR}(K_L \to \mu^+ \mu^-)$. Observables sensitive to $d_i \to d_j \overline{\nu} \nu$ transitions are also included, the most important being $\text{BR}(K^+ \to \pi^+ \nu_\mu \overline{\nu_\mu})$, remembering that, for simplicity, we set the imaginary parts of Wilson coefficients to zero and so have no sensitivity to {\em e.g.} $\text{BR}(K_L \to \pi^0 \nu_\mu \overline{\nu_\mu})$. 

We overlay the constraints coming from finite naturalness of the Higgs mass, for leptoquark masses equal to 4, 5, and 5.5 TeV. For each mass, the region above the black line has a tuning $\Delta > 1$. For heavy leptoquark masses, we can see that the Higgs mass becomes destablized (in the sense that $\Delta$ exceeds one) when the new physics is aligned with the third family, {\em i.e.} when $\alpha_3 \gg \alpha_2$. Specifically, we find that a $U(2)$-like flavour structure, here defined by \(\alpha_2 / \alpha_3 \in [0.2,5] \V{ts}\) as indicated by the hatched red regions in Figs.~\ref{fig:a3a2}, \ref{fig:MLQ}, \& \ref{fig:ROFV}, becomes disfavoured by finite naturalness for leptoquark masses greater than 5 TeV or so, as can be seen also in Fig.~\ref{fig:MLQ}.

The decays $B\to K^{(*)} \nu\bar{\nu}$, while in principal important, are in fact constrained by precisely the same Wilson coefficient, $\Delta C_9^\mu (=-\Delta C_{10}^\mu)$ that mediates the $B$ anomalies. Again, this is a consequence of our gauging $L_\mu - L_\tau$, which means the leptoquark only has decays to muon neutrinos.
The theory prediction for 
$R_{\nu\nu}^{(\ast)} \equiv \text{BR}\left(B \to K^{(\ast)}\nu\nu \right)/\,\text{BR}\left(B \to K^{(\ast)}\nu\nu\right)_{\text{SM}}$
is~\cite{Buras:2014fpa}
\be \label{eq:Rnunu}
R_{\nu\nu}^{(\ast)} \approx 1 + \frac{1}{3}\frac{\text{Re}\{\Delta C_9^{\mu} \}}{C_L^{\text{SM}}},
\ee
where $C_L^{\text{SM}}=-6.38\pm 0.06$~\cite{Altmannshofer:2009ma}. The tighter experimental bound comes from the Belle measurement $R_{\nu\nu}^\ast < 2.7$ at 2$\sigma$~\cite{Grygier:2017tzo}, which gives only a very weak constraint
$M_{S_3}/\sqrt{\alpha_3 \alpha_2^*} \geq 4.4 \text{~TeV}$,
which is safely satisfied for the entire 2$\sigma$ range that fits the $B$ anomaly data. Turning this around, one might rather interpret the BSM correction in Eq. (\ref{eq:Rnunu}) as a prediction of our model; assuming the best fit value of $\Delta C_9^\mu$, we expect $R_{\nu\nu}^{(\ast)} \approx 1.026$, {\em i.e.} only a 2.6\% correction to the SM prediction (which is significantly smaller than for typical leptoquark models, which often feature large branching ratios to tau neutrinos). Indeed, taking only the 2$\sigma$ range of the fit to $B$ anomaly data, we expect the correction to $R_{\nu\nu}^{(\ast)}$ to be no greater than 3.6\%.

Finally, we checked a number of other observables which gave only very weak constraints. This includes constraints from $b\bar{b}\to \mu^+\mu^-$, which implies $M_{S_3}/|\alpha_3| \geq 1.8 \text{~TeV}$; constraints from Higgs production and decay as well as the oblique parameters $S$ and $T$, which are modified by the loop-induced coupling $\lambda^\prime_{HS}$, but which give even weaker constraints; and constraints from the lifetime ratio \(\tauBs/\tauBd\).

\section{Discussion}

In this paper we put forward a scenario in which the SM gauge group is extended by a $U(1)_X$ gauge symmetry, where $X=L_\mu-L_\tau$ on the SM fermions, and its matter content is augmented by a scalar leptoquark that has charge $-1$ under $U(1)_X$. By charging the leptoquark under $L_\mu - L_\tau$ we resolve two conceptual problems with $S_3$ models, both of which result from the fact that an uncharged leptoquark extension does not respect the SM's accidental symmetries. These accidental symmetries, specifically baryon number and the three individual charged lepton numbers, are very good symmetries of Nature, at least for energy scales probed by contemporary colliders.

Specifically, charging the $S_3$ leptoquark bans the presence of renormalisable diquark operators in the lagrangian, and in fact pushes back baryon number-violating effects to dimension-6 in the effective field theory. Thus, we expect the proton to decay no faster in our model than in the Standard Model effective field theory. On the leptonic side, by gauging $L_\mu - L_\tau$ there is no danger of lepton flavour violation, and the $S_3$-mediated FCNCs appear exclusively in muons, which is in good agreement with global fits to the $b\to s \ell\ell$ anomaly data.

By gauging $L_\mu - L_\tau$, which we assume is spontaneously broken somewhere above the weak scale, a new heavy gauge boson and heavy scalar appear in the spectrum of the theory. This pair of particles play no role in mediating the $B$ anomalies, and we show that they may be hidden away at very high mass scales without destabilizing either the Higgs mass or the leptoquark mass through loop corrections. However, we do find that the Higgs mass inevitably receives significant loop corrections due to the leptoquark itself, whose couplings to the top quark must be big enough to explain the $B$ anomalies. This is the case for any $S_3$ leptoquark that mediates the anomalies in $b \to s \ell \ell$, a fact that seems to have escaped attention in previous literature. By requiring the Higgs mass not be fine-tuned against its finite one-loop corrections, we obtain bounds on the mass and couplings of the leptoquark.

Because the Higgs mass correction is largest due to a loop involving top-quarks, this finite-naturalness argument requires the coupling $\alpha_3$ of the leptoquark to the third family quark doublet to be not too big. Intriguingly, for leptoquark masses greater than about 5.2 TeV or so, one struggles to accommodate a $U(2)$-like flavour structure 
without introducing significant tuning of the Higgs mass. For such masses, naturalness therefore favours a larger degree of `strange-alignment' in the leptoquark couplings.
For leptoquarks heavier than 5.8 TeV, the Higgs mass requires significant tuning regardless of the flavour structure, because of an unavoidable electroweak two-loop contribution.

To summarise the results of our finite naturalness study,
there is a wide (but narrowing) window of $U(2)$-compatible $S_3$ leptoquark masses, $M_{S_3} \in [1.7,5.2]$ TeV, which are heavy enough to evade direct searches at the LHC but light enough to remain (finitely) natural. It will be interesting to chart the fate of this $U(2)$-compatible mass window in the near future, as LHC searches push $M_{S_3}$ higher, and as correlated flavour observables are measured with increasingly good precision.

\acknowledgments{We are very grateful to Wolfgang Altmannshofer for useful discussions and participation in the early stages of this project, and for comments on the manuscript. We also thank Luca di Luzio for a helpful discussion regarding beta functions.
JD is supported by the STFC consolidated grant ST/P000681/1, and thanks other members of the Cambridge Pheno Working Group for discussions. MK was supported by MIUR (Italy) under a contract PRIN 2015P5SBHT and by INFN Sezione di Roma La Sapienza and partially supported by the ERC-2010 DaMESyFla Grant Agreement Number: 267985.}

\appendix

\section{Loop induced couplings}
\label{app:loop_induced}

To estimate the size of the radiatively-generated coupling between the leptoquarks and the Higgs (or the \(\Phi\) field), we take the one-loop beta functions \(\beta \equiv d \lambda / d \ln \mu^2\), and make the estimate
\begin{equation}
\delta\lambda (\mu) \approx \left(\text{lim}_{\lambda \to 0}\; \beta_\lambda\right) \ln \frac{\mu^2}{M^2},
\end{equation}
where \(M\) is the mass of the virtual particle in the loop, and we take the tree-level scalar coupling to zero as part of our assumption that the one-loop effect is dominant.

Computing the $\beta$ functions for scalar quartic coupling are standard computations \cite{Cheng:1973nv,Machacek:1984zw,Schienbein:2018fsw}, and we extract the results for $\lambda_{HS}$, $\lambda^\prime_{HS}$, and $\lambda_{\Phi S}$ as follows:
\begin{align}
\text{lim}_{\lambdaHSprime, g \to 0}\;\beta_{\lambdaHSprime} &= -\frac{G_F m_t^2 |\sum_q \alpha_q V_{tq}|^2}{2 \sqrt{2} \pi^2}, \\
\text{lim}_{\lambda_{HS}, g \to 0}\;\beta_{\lambda_{HS}} &\approx 0, \\
\intertext{(where for these first two we have dropped the terms in the beta function proportional to the electroweak coupling \(g\) to avoid double counting against the 2-loop result Eq.~\ref{eq:higgs_2loop})}
\text{lim}_{\lambda_{\Phi S} \to 0}\;\beta_{\lambda_{\Phi S}} &= -\frac{g_X^4 \hat{Q}_\Phi^2}{4 \pi^2} + \BigO{\frac{M_{S_3}^4}{M_X^4}}. \label{eq:beta_PhiS}
\end{align}
The contributions to $\lambda_{HS}$ occur due to diagrams with bottom quarks running in loops, and so $\beta_{\lambda_{HS}}$ is suppressed with respect to  $\beta_{\lambda^\prime_{HS}}$ by a factor of \(m_b^2 / m_t^2 \sim 10^{-3}\), which we safely neglect throughout.
Most importantly for the discussion in the main text, we thus estimate
\begin{equation}
\lambdaHSprime (\mu) \approx -\frac{G_F m_t^2 |\sum_q \alpha_q V_{tq}|^2}{2 \sqrt{2} \pi^2}\ln \frac{\mu^2}{m_t^2} \approx -0.013 |\alpha_3 + V_{ts} \alpha_2|^2 \ln \frac{\mu^2}{m_t^2}
\end{equation}
and $\lambdaHS, \lambdaPhiS \approx 0$.

\section{Scalar one-loop mass corrections}
\label{app:mass_corrections}

Firstly, define the following two loop functions,
\begin{align}
A_0 (M) &= M^2 \left( 1 + \ln \frac{\mu^2}{M^2}\right) \equiv \texttt{PVA[0,M]} \,, \\
B_0 (M_1, M_2, M_3) &=  2 + \Lambda(M_1^2, M_2, M_3) - \frac{(M_1^2 + M_2^2 - M_3^2)\ln \frac{M_2^2}{M_3^2}}{2 M_1^2} + \ln \frac{\mu^2}{M_3^2} \\
&\equiv \texttt{PVB[0,0,M1\^{}2,M2,M3]},
\end{align}
where the notation \texttt{PVA} and \texttt{PVB} are borrowed from \texttt{Package-X} syntax~\cite{PackageX,Patel:2015tea,Patel:2016fam}, where
\begin{equation}
\Lambda(M_1^2, M_2, M_3) = \frac{\sqrt{\lambda(M_1^2,M_2^2,M_3^2)}}{M_1^2} \ln \left( \frac{\sqrt{\lambda(M_1^2,M_2^2,M_3^2)}-M_1^2+M_2^2+M_3^2}{2 M_2 M_3}\right),
\end{equation}
and where \(\lambda\) is the Källén function
\begin{equation}
\lambda (a, b, c) = a^2 + b^2 + c^2 - 2ab - 2ac -2bc.
\end{equation}
These loop functions appear in the various mass corrections we compute in the rest of this Appendix.

\subsection{Higgs mass corrections}

There are three Feynman diagrams that contribute to the Higgs mass at one-loop, as shown in Fig.~\ref{fig:DeltaMH} (in the main text), all of which involve a leptoquark running in a loop. Considering the diagrams in order from left to right in Fig.~\ref{fig:DeltaMH}, there is firstly a diagram in which the leptoquark loop attaches to the Higgs propagator at a four-point vertex, giving
\begin{equation}
\delta M_H^2 =\frac{9}{32 \pi^2} (\lambdaHS + \lambdaHSprime) A_0 (M_{S_3}).
\end{equation}
Secondly, there is a tadpole diagram giving
\begin{equation}
\delta M_H^2 =\frac{-27}{32 \pi^2} (\lambdaHS + \lambdaHSprime) A_0 (M_{S_3}).
\end{equation}
Finally, there is a diagram involving two insertions of the Higgs-$S_3$-$S_3$ vertex, which gives
\begin{equation}
\delta M_H^2 = \frac{3 v^2}{16 \pi^2} (3\lambdaHS^2 + 6 \lambdaHS \lambdaHSprime + 5 \lambda^{\prime2}_{HS}) B_0(M_H, M_{S_3}, M_{S_3}),
\end{equation}
In the limit of interest, $M_{S_3} \gg M_H$, the first two diagrams give the leading contributions and we neglect the third, resulting in Eq.~(\ref{eq:higgs_mass_correction}) from the main text.

\subsection{Finite naturalness of the leptoquark mass} \label{app:LQ2}

In this Appendix we discuss the finite naturalness of the leptoquark mass, which receives a multitude of one-loop corrections involving both SM particles and the heavy fields associated with the $U(1)_X$ symmetry. In the $U(1)_X$ decoupling limit (\ref{eq:decoupling}) that we explore in the main text, we find that the relative size of these loop corrections is small, giving only weak constraints on the parameter space of our charged leptoquark model. These constraints do not appear in any of Figs.~\ref{fig:a3a2},~\ref{fig:MLQ}, or~\ref{fig:ROFV}, where we plot only the finite naturalness contours coming from the Higgs mass tuning.

Before we discuss radiative corrections, we remark that there are several `tree-level' contributions to the leptoquark mass encoded in the lagrangian (\ref{eq:LQ-quark-lepton}). In addition to the bare mass term, the potential terms $-\lambda_{HS}|H|^2|S_3|^2$ and $-\lambda^\prime_{HS}|H^\dagger \sigma_a S_3^a|^2$ give leptoquark mass contributions once the Higgs is expanded around its vev. Moreover, the contributions from the $\lambda^\prime_{HS}$ term are different for each of the three $SU(2)$ components of $S_3$, giving a mass splitting
\begin{equation}
\Delta_M = M_{S_3^{1/3}} - M_{S_3^{4/3}} = M_{S_3^{-2/3}} - M_{S_3^{1/3}} = \frac{\lambda'_{HS}}{4} \frac{v^2}{M_{S_3^{4/3}}} + \BigO{\frac{v^4}{M_{S_3}^3}},
\end{equation}
which is small for the multi-TeV mass leptoquarks of interest here.
There is also a tree-level mass contribution from the potential term $-\lambda_{\Phi S}|\Phi|^2|S_3|^2$ upon expanding $\Phi$ about its vev $v_\Phi$, which could be large given that we expect $U(1)_X$ to be broken at some higher energy scale. Indeed the coupling $\lambda_{\Phi S}$, like the couplings $\lambda_{HS}^{(\prime)}$ discussed above, is generated at one-loop. From Eq. (\ref{eq:beta_PhiS})
we expect this mass contribution to  scale like $\delta M_{S_3^{4/3}}^2 \sim g_X^2 M_X^2/4\pi^2$, which is comparable to the one-loop corrections that we discuss next.

\begin{figure}
\includegraphics[width=0.8\textwidth]{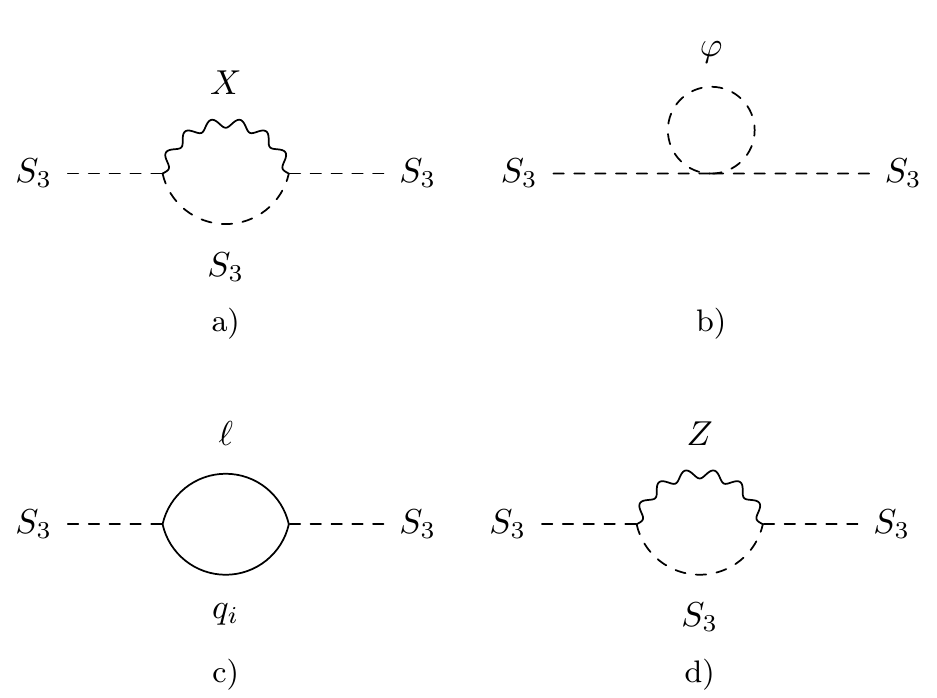}
\caption{Example one-loop Feynman diagrams contributing to radiative corrections to the leptoquark mass, from a) the \(X\) gauge boson, b) the \(\Phi\) scalar, c) SM fermions, and d) EW gauge bosons.}
\label{fig:DeltaMS3}
\end{figure}
We now discuss radiative corrections, and finite naturalness of the leptoquark mass. Within our assumed mass hierarchy, the largest corrections are na\"ively those proportional to \(M_\Phi\) and \(M_X\), coming from these heavier particles running in loops.
Neglecting small electroweak-nonuniversal contributions, the dominant mass corrections are (see Figs.~\ref{fig:DeltaMS3}a and~\ref{fig:DeltaMS3}b for the responsible Feynman diagrams)
\begin{equation} \label{eq:LQ_mass_X}
(\delta M_{S_3}^2)_X = 
\begin{aligned}[t]
&\frac{g_X^2 M_X^2}{4\pi^2} - \frac{\lambdaPhiS M_X^2}{4\pi^2} \frac{M_X^2}{M_\Phi^2} - \frac{\lambdaPhiS M_\Phi^2}{8 \pi^2} \\
+ &\left( \frac{g_X^2 M_X^2}{2\pi^2} - \frac{\lambdaPhiS M_X^2}{2\pi^2} \frac{M_X^2}{M_\Phi^2} \right) \ln \frac{\mu^2}{M_X^2} - \frac{\lambdaPhiS M_\Phi^2}{8 \pi^2} \ln \frac{\mu^2}{M_\Phi^2}+ \dots,
\end{aligned}
\end{equation}
where the subscript `$X$' indicates that these corrections are due to particles in the $U(1)_X$ sector. In the decoupling limit specified by Eq. (\ref{eq:decoupling}) these contributions become vanishingly small.

In that limit, the leading one-loop corrections to the leptoquark masses are due to various SM particles running in the loop. Again ignoring the small electroweak-nonuniversal corrections,\footnote{We here also ignore terms arising from the $\lambda_{HS}^{(\prime)}$ couplings, which we assume are suppressed by a further loop factor (as in \S \ref{sec:Hmass}).} we have (see Fig.~\ref{fig:DeltaMS3}c)
\begin{equation}
(\delta M_{S_3}^2)_\text{fermion} = -\frac{M_{S_3}^2}{8\pi^2}\left(\sum_i |\alpha_i|^2\right) \left(2 + \ln \frac{\mu^2}{M_{S_3}^2} \right)\, ,
\end{equation}
coming from the quark-lepton loops, 
which gives only very weak finite naturalness bounds
\begin{equation}
\sum_i |\alpha_i|^2 \lesssim
\begin{cases}
	40 \Delta &\text{ for } \mu = M_{S_3} \\
	\Delta &\text{ for } \mu = 10^{16} M_{S_3} \sim M_\text{Planck}.
\end{cases}
\end{equation}
The contribution from EW gauge bosons is (Fig.~\ref{fig:DeltaMS3}d)
\begin{equation}
\begin{aligned}
(\delta M_{S_3}^2)_\text{EW} &= -\frac{M_{S_3}^2}{144\pi^2}\left( 16 e^2 + 9 g^2 \left\{ 1 + \frac{(s_W^2 - 3 c_W^2)^2}{9c_W^2} \right\} \right) \left( 7 + 3 \ln \frac{\mu^2}{M_{S_3}^2} \right) \\ &= 
\begin{cases}
	-0.03 M_{S_3}^2 &\text{ for } \mu = M_{S_3} \\
	-1.2 M_{S_3}^2 &\text{ for } \mu = 10^{16} M_{S_3} \sim M_\text{Planck}.
\end{cases}
\end{aligned}
\end{equation}
Thus, if we take the ``Planck scale'' condition then a degree of fine tuning is unavoidable (assuming there is no cancellation between different mass corrections). This follows simply from the existence of an $S_3$ leptoquark charged under the SM gauge group, even without any fermion couplings.

\bibliographystyle{utphys}
\bibliography{articles}
\end{document}